\documentclass[twocolumn]{aastex631}

\usepackage{graphicx}
\usepackage{amssymb,amsmath}
\usepackage{aas_macros}
\usepackage[T1]{fontenc}
\usepackage{bm}
\usepackage{xcolor}
\usepackage{multirow}

\newcommand{\dd}{\ensuremath{\mathrm{d}}}%
\newcommand{\vct}[1]{\ensuremath{\bm{#1}}}%
\newcommand{\uvct}[1]{\hat{\bm #1}}
\newcommand*{\ii}{\imath}
\newcommand{\Lmax}{\ensuremath{L_{\text{max}}}}%
\newcommand{\Lc}{\ensuremath{L_{\text{c}}}}%
\newcommand{\rg}{\ensuremath{r_{\text{g}}}}%
\newcommand{\Brms}{\ensuremath{B_{\text{rms}}}}%
\newcommand{\taus}{\ensuremath{\tau_{\text{s}}}}%
\newcommand{\tauc}{\ensuremath{\tau_{\text{c}}}}%
\newcommand{\thetaeff}{\ensuremath{\theta_{\mathrm{eff}}}}%
\newcommand{\sigmaeff}{\ensuremath{\sigma_{\mathrm{eff}}}}%
\DeclareMathOperator{\csch}{csch}%

\begin{document}

\title{Diffusion of relativistic charged particles and field lines in isotropic turbulence: I. Numerical simulations}

\author[0000-0002-0454-6823]{Marco Kuhlen}
\affiliation{Institute for Theoretical Particle Physics and Cosmology (TTK), RWTH Aachen University, 52056 Aachen, Germany}
\email{marco.kuhlen@rwth-aachen.de}

\author[0000-0002-2197-3421]{Philipp Mertsch}
\affiliation{Institute for Theoretical Particle Physics and Cosmology (TTK), RWTH Aachen University, 52056 Aachen, Germany}
\correspondingauthor{Philipp Mertsch} 
\email{pmertsch@physik.rwth-aachen.de}

\author[0000-0002-5611-095X]{Vo Hong Minh Phan}
\affiliation{Sorbonne Université, Observatoire de Paris, PSL Research University, LERMA, CNRS UMR 8112, 75005 Paris, France}
\affiliation{Institute for Theoretical Particle Physics and Cosmology (TTK), RWTH Aachen University, 52056 Aachen, Germany}
\email{vhmphan@physik.rwth-aachen.de}

\preprint{TTK-22-35}

\date{\today}

\begin{abstract}
The transport of non-thermal particles across a large-scale magnetic field in the presence of magnetised turbulence has been a long-standing issue in high-energy astrophysics. Of particular interest is the dependence of the parallel and perpendicular mean free paths $\lambda_{\parallel}$ and $\lambda_{\perp}$ on rigidity $\mathcal{R}$. We have revisited this important issue with a view to applications from the transport of Galactic cosmic rays to acceleration at astrophysical shocks. We have run test particle simulations of cosmic ray transport in synthetic, isotropic Kolmogorov turbulence at unprecedentedly low reduced rigidities $\rg/\Lc \simeq 10^{-4}$, corresponding to $\mathcal{R} \simeq 10 \, \text{TV}$ for a turbulent magnetic field of $\Brms = 4 \, \mu\text{G}$ and correlation length $\Lc = 30 \, \text{pc}$. Extracting the (asymptotic) parallel and perpendicular mean free paths $\lambda_{\parallel}$ and $\lambda_{\perp}$, we have found $\lambda_{\parallel} \propto (\rg/\Lc)^{1/3}$ as expected for a Kolmogorov turbulence spectrum. In contrast, $\lambda_{\perp}$ has a faster dependence on $\rg/\Lc$ for $10^{-2} \lesssim \rg/\Lc \lesssim 1$, but for $\rg/\Lc \ll 10^{-2}$, also $\lambda_{\perp} \propto (\rg/\Lc)^{1/3}$. Our results have important implications for the transport of Galactic cosmic rays. 
\end{abstract}



\section{Introduction\label{sec:introduction}}

There is overwhelming observational evidence that the transport of high-energy, non-thermal particles in the Galaxy is predominantly diffusive: For Galactic cosmic rays (CRs), the grammage, that is the inferred amount of matter that CRs experience on average, implies that CRs cross the gaseous disk of the Galaxy many times between acceleration and observation. The small anisotropies in the observed arrival directions of CRs also point to an efficient randomisation of CR directions. (See, e.g.\ \citet{2019IJMPD..2830022G} for a review on Galactic CRs.) For collisionless plasmas like CRs, the interactions with turbulent magnetic fields act as the sole agent of such randomisation. 

The first attempt at clarifying the interaction of waves and particles was the so-called Quasi-Linear Theory (QLT), a perturbative approach that considers a small turbulent magnetic field $\vct{\delta B}$ on top of the regular background field $\vct{B}_0$~\citep{1966ApJ...146..480J,1966PhFl....9.2377K,1967PhFl...10.2620H,1970ApJ...162.1049H}. The scattering rate is estimated by approximating the force experienced by particles on unperturbed trajectories, that is solutions of the equation of motion in the background field $\vct{B}_0$ only. Magnetic turbulence is oftentimes modelled as a spectrum of different waves. There exists a few different types of interactions between the turbulent field and the charged particles including transit-time damping \citep{schlickeiser1998,xu2018}, interactions with magnetic mirrors \citep{lazarian2021}, and gyro-resonant interactions within which particles with a gyroradius $\rg$ are scattered by modes with a wavenumber $k \sim 1/\rg$. Under this gyro-resonant condition, a coherent force accumulates over many gyrations resulting in scattering of the particle's pitch-angle, i.e.\ the angle between the particle momentum and $\vct{B}_0$. Scattering by an ensemble of wave-like perturbations, a particle's pitch-angle becomes diffuse and once particles have reversed direction, the spatial transport along the background field also becomes diffusive~\citep{1970ApJ...162.1049H}. Due to the resonant nature of the interactions, the mean free path $\lambda_{\parallel}$ will depend on particle rigidity~\footnote{Rigidity $\mathcal{R}$ is defined as the ratio of particle momentum $p$ and charge $q$, that is $\mathcal{R} \equiv p c / q$ with the speed of light $c$.} and the turbulence power spectrum. Here, we introduce the reduced rigidity $\rg{}/\Lc$ where $\rg{} = p c / (e \Brms) = \mathcal{R} / \Brms$ is the relativistic gyroradius, $\Lc$ denotes the correlation length, $p$ is the momentum and $e$ the charge of the particle. Generically, for a $k^{-5/3}$ Kolmogorov power spectrum~\citep{Kolmogorov:1931} for instance, $\lambda_{\parallel} \propto (\rg/\Lc)^{1/3}$. 

Applying QLT to slab turbulence, where the wavevector $\vct{k} \parallel \vct{B}_0$, the pitch-angle diffusion coefficient vanishes for pitch-angles close to $90^{\circ}$. This is known as the $90^{\circ}$ problem~\citep{1975RvGSP..13..547V}. The parallel mean free path, $\lambda_{\parallel}$, which is proportional to the inverse of the pitch-angle diffusion coefficient, however, remains finite if the turbulence spectrum is not too steep~\citep{2009ASSL..362.....S}. For isotropic turbulence, instead, $\lambda_{\parallel}$ diverges~\citep{2006JPhG...32..809T}. Various modifications of QLT have been suggested to cure this problem. (See \citet{2020SSRv..216...23S} for a recent review.) The most promising suggestions are perhaps dynamical turbulence~\citep{2006JPhG...32.1045T}, non-linear theories~\citep{2009ASSL..362.....S}, and interactions with magnetic mirrors \citep{cesarsky1973,felice2001}. Both dynamical turbulence and non-linear theories lead to a broadening of the resonance condition and can hence cure the vanishing of the pitch-angle diffusion coefficient. Interactions with magnetic mirrors, on the other hand, introduce a critical value of the pitch-angle below which the effect of magnetic mirroring due to modes of long wavelength becomes more dominant than gyro-resonant interactions and, thus, allow for a finite value of the pitch-angle diffusion coefficient around $90^{\circ}$. Transport in the perpendicular direction, however, is a different matter altogether. It is believed that it depends both on the transport of particles along field lines as well as the transport of the field lines itself. For the composite turbulence models popular in the space science communities, various theories have been developped~\citep{BAM-model,2003ApJ...590L..53M,2004ApJ...616..617S}. See again \citet{2020SSRv..216...23S} for an account of successes and failures of such theories. 

In simulations, the equations of motions are solved for a large number of particles in individual realisations of the turbulent magnetic field $\vct{\delta B}$. This field is either the result of MHD simulations~\citep{2011ApJ...728...60B,2013ApJ...779..140X,2016A&A...588A..73C,2020arXiv201000699S} or has been created as ``synthetic turbulence'', that is according to a heuristic turbulence model (e.g. isotropic turbulence with a power law power spectrum)~\citep{1994ApJ...430L.137G,1997A&A...326..793M,1998A&A...337..558M,1999ApJ...520..204G,2001PhRvD..65b3002C,2002GeoRL..29.1048Q,2007JCAP...06..027D,2015JGRA..120.4095H,2016MNRAS.459.3395P,2018JCAP...07..051G,2020MNRAS.498.5051R}. While MHD turbulence is more realistic, its dynamical range is limited; conversely, synthetic turbulence offers a larger dynamical range, but it does not capture the full dynamics of MHD turbulence. However, even in synthetic turbulence, the simulations have not been performed at GV or TV rigidities for which the most precise direct observations of CRs are available. Extrapolations from larger rigidities are of limited value due to the possibility that the transport regime changes with rigidity. Various features observed in the spectra of TV CRs could be taken as hints for such transitions. See \citet{2020Ap&SS.365..135M} for a review of test particle simulations. 

In the absence of a reliable theory and of simulations at the relevant rigidities, the simplest expectation is for the parallel and perpendicular mean free path to scale with rigidity in the same way while the dependence on $\delta B^2 / B_0^2$ should be the opposite~\citep{1980gbs..bookQ....M,yan2008}, 
\begin{align}
\lambda_{\parallel} & \propto \left(\frac{\rg}{\Lc}\right)^{1/3} \left( \frac{\delta B^2}{B_0^2} \right)^{-1} \, , \label{eqn:lambda_par_generic} \\
\lambda_{\perp} & \propto \left(\frac{\rg}{\Lc}\right)^{1/3} \left( \frac{\delta B^2}{B_0^2} \right) \, . \label{eqn:lambda_perp_generic}
\end{align}
Numerical simulations of isotropic turbulence, however, show a different scaling of $\lambda_{\parallel}$ and $\lambda_{\perp}$. While $\lambda_{\parallel}$ is in agreement with the $(\rg/\Lc)^{1/3}$ behaviour generically expected for gyro-resonant interactions with Kolmogorov turbulence, the scaling of $\lambda_{\perp}$ with $\rg/\Lc$ is faster. This behaviour had been indicated already in the simulations by De Marco \textit{et al.}~\citep{2007JCAP...06..027D}. The simulations by \citet{2016MNRAS.457.3975S} and \citet{2018JCAP...07..051G} show a similar trend, even though the respective authors refrain from a verdict on the reliability of this result. More recently, \citet{2020PhRvD.102j3016D} investigated this behaviour in more detail, finding a ratio $\lambda_{\perp} / \lambda_{\parallel} \propto (\rg/\Lc)^{\Delta s}$ with $\Delta s \simeq 0.2$. We also note that at the reduced rigidities where simulation results are available, the scaling of $\lambda_{\perp}$ with $(\delta B / B_0)$ indicated in eq.~\eqref{eqn:lambda_par_generic} is not observed either, see for instance Fig.~6 of \citet{2020Ap&SS.365..135M}. Simulational results of particle propagation in MHD turbulence, in fact, points towards $\lambda_\perp/\lambda_\parallel\sim \delta B^4/B_0^4$ \citep{xu2013,maiti2022}. These studies are, however, on particles with rather large reduced rigidities $\rg/\Lc$ and also for $\delta B/B_0<1$ which might not be very straightforward to apply in the context of Galactic CR transport. So far, neither the scaling with rigidity nor with $(\delta B / B_0)$ has been understood. 

Note that for a correlation length of $30 \, \text{pc}$ and $\Brms = 4 \, \mu\text{G}$, reduced rigidities $\rg{}/\Lc \simeq 10^{-2} \mathellipsis 10^{-1}$ correspond to rigidities $\mathcal{R} \simeq (1 \mathellipsis 10) \, \text{PV}$. If extrapolated to GV rigidities, this difference could have important phenomenological consequences. Diffuse emission at GeV energies is sensitive to the spectra and spatial distribution of CR nuclei and leptons of GV and TV rigidities. Some of the deviations between isotropic diffusion models and observations have been suggested to be due to a different rigidity-dependence of $\lambda_{\parallel}$ and $\lambda_{\perp}$~\citep{2012PhRvL.108u1102E,2017JCAP...10..019C,2020MNRAS.498.5051R}. Anisotropic diffusion might also play an important role in the confinement of CRs in different parts of the Galaxy and in the overall escape at large rigidities~\citep{2018JCAP...07..051G}. 

We have revisited the important question on the nature of perpendicular transport with a particular focus on the rigidity-dependence of the perpendicular mean free path $\lambda_{\perp}$ in isotropic turbulence. To this end, we have run a large suite of numerical test-particle simulations reaching unprecedentedly low rigidities, $\rg{}/\Lc \simeq 10^{-4}$ which for $\Lc = 30 \, \text{pc}$ and $\Brms = 4 \, \mu\text{G}$ corresponds to rigidities of $\mathcal{R} \sim 10 \, \text{TV}$. These small rigidities have been made possible by the use of graphics processing units (GPUs) for solving the equations of motions. (See \citet{2016NewA...45....1T} for an earlier use of GPUs in test particle simulations.) We have found that the perpendicular mean free path $\lambda_{\perp}$ scales differently than the parallel one for $10^{-2} \lesssim \rg/\Lc \lesssim 1$, but for $\rg/\Lc \ll 10^{-2}$ the same rigidity-dependence is recovered. If the transition from the standard $(\rg/\Lc)^{1/3}$-dependence at low values of $\rg/\Lc$ to the faster dependence on $\rg/\Lc$ at intermediate values of $\rg/\Lc$ depended on $(\delta B^2 / B_0^2)$, this could explain why at intermediate values of $\rg/\Lc$, a $(\delta B^2 / B_0^2)$-dependence different from eq.~\eqref{eqn:lambda_perp_generic} was observed. 

The outline of this paper is as follows. 
In Sec.~\ref{sec:methods}, we explain the preparation of an isotropic, turbulent magnetic field on a computer, describe the test particle simulations and introduce the various correlation functions and diffusion coefficients we determined from the simulation results. Our results are presented and discussed in Sec.~\ref{sec:results}.
We put our results into context in Sec.~\ref{sec:discussion}, offering some thoughts on the phenomenological consequences of our findings, and conclude in Sec.~\ref{sec:conclusion}.

\section{Methods\label{sec:methods}}

\subsection{Test particle simulations}
\label{sec:turbulence}

In order to study the perpendicular transport of cosmic rays in the presence of a turbulent magnetic field, we will be making use of test particle simulations, meaning that the back-reaction of cosmic rays onto the magnetic field is neglected (see \citet{2020Ap&SS.365..135M} for a review). In this limit the magnetic field is defined once, before a set of test particle trajectories is computed by solving the Newton-Lorentz equation
\begin{equation}
  \frac{\mathrm{d}}{\mathrm{d}t}\vct{p} = q\left[\vct{E}(\vct{r},t)+\frac{\vct{v}}{c}\times\vct{B}(\vct{r},t)\right] \, ,
  \label{eqn:Newton-Lorentz}
\end{equation}
where $q$ is the particle charge, $\vct{v}$ and $\vct{p}$ are the velocity and momentum vectors, and $\vct{E}$ and $\vct{B}$ represent the electric and magnetic vector fields. We will be assuming magnetostatic turbulence throughout, meaning that $\partial \vct{B}/\partial t = 0$. The electric field is neglected since the high mobility of charges in typical astrophysical plasmas is effectively preventing the presence of any large scale electric field $\vct{E}_0$. On scales larger than the Debye length the electric charges effectively shield any charge overdensity producing an electric potential~\citep{Debye}. While a small scale electric field $\vct{\delta E}$ could in principle be present, it follows from Faraday's induction law that for Alfvén-wave type magnetic field perturbations, $|\vct{\delta E}| \sim (v_A/c)|\vct{\delta B}|$ with $v_A$ denoting the Alfvén velocity. Since in most astrophysical environments the Alfvén velocity is much smaller than the speed of light, i.e. $v_A/c\simeq 10^{-5}-10^{-3}$ for typical interstellar medium conditions~\citep{2001RvMP...73.1031F}, the small scale electric field can also be neglected here. Note that in the absence of an electric field, the particle energy is conserved. This has to hold for the test particle trajectories in the simulations also.

The computed test particle trajectories can then be used to infer parameters of the transport model, such as the diffusion coefficient~\citep{1994ApJ...430L.137G,1997A&A...326..793M,1998A&A...337..558M,1999ApJ...520..204G,2001PhRvD..65b3002C,2002GeoRL..29.1048Q,2007JCAP...06..027D,2018JCAP...07..051G,2020MNRAS.498.5051R}, predict spectra or large-scale anisotropies~\citep{Giacinti:2014xya,Savchenko:2015dha,genolini2021}, or investigate effects beyond the standard picture of cosmic ray transport such as the generation of small-scale anisotropies in the arrival directions~\citep{giacinti2012,2015ApJ...815L...2A,lopez-barquero2016,2017PrPNP..94..184A,2019JCAP...11..048M,2022ApJ...927..110K}.

The total magnetic field for these simulations is modelled as the superposition of a large-scale, coherent field $\vct{B}_0$ and a small-scale, turbulent field $\vct{\delta B}$. In particular, we will consider synthetic isotropic Kolmogorov turbulence with the two-point correlation function of the following form (see \citet{2020Ap&SS.365..135M} for more detailed discussions on some common synthetic turbulence models)
\begin{equation}
  \begin{aligned}
    P_{ij}(\vct{k}) &\equiv \int \mathrm{d}^3 r \, e^{\ii \vct{k}\cdot\vct{r}}\langle\delta B_i(\vct{r}_0)\delta B_j(\vct{r}_0+\vct{r})\rangle\\
    &= \frac{g(k)}{k^2} \left(\delta_{ij} - \frac{k_ik_j}{k^2}\right)
  \end{aligned}
\end{equation}
where the one-dimensional power spectrum $g(k)$ is a simple power law, with the spectral index for Kolmogorov turbulence $\gamma = 5/3$ ~\citep{1941DoSSR..30..301K}, given as
\begin{equation}
g(k) = \left\{ \begin{array}{ll} \delta B^2 k^{-\gamma} \frac{\vphantom{\frac{a}{b_{a_a}}}(\gamma-1)\left(\frac{2\pi}{L_\text{max}}\right)^{\gamma-1}}{\vphantom{\frac{b^{a^a}}{a}} 1-\frac{L_\text{min}}{L_\text{max}}} & \text{for } \frac{2\pi}{L_\text{max}}\leq k \leq \frac{2\pi}{L_\text{min}} \, , \\
0 & \text{otherwise.} \end{array} \right. 
\end{equation} 

We characterise the overall strength of the turbulent component with respect to the total rms B-field through the turbulence level $\eta$,
\begin{equation}
\eta \equiv \frac{\delta B^2}{B_0^2 + \delta B^2} \, .
\label{eqn:def_eta}
\end{equation}

Given the one-dimensional power spectrum, one can also define the correlation length of the magnetic field $\Lc\equiv\int_{-\infty}^{\infty}\mathrm{d}L~\langle\vct{\delta B}(\vct{r}_0)\cdot\vct{\delta B}(\vct{r}_0+\vct{\Delta r}(L))\rangle/\delta B^2$ where $(\vct{r}_0+\vct{\Delta r}(L))$ is displaced with respect to $\vct{r}_0$ by a distance $L$ along a fixed direction~\citep{2002JHEP...03..045H}. For a power law magnetic power spectrum, the correlation length evaluates to 
\begin{equation}
  \Lc = \frac{1}{2} L_\text{max} \frac{\gamma-1}{\gamma} \frac{1-(L_\text{min}/L_\text{max})^{\gamma}}{1-(L_\text{min}/L_\text{max})^{\gamma-1}} \, ,
\end{equation}
which gives $\Lc \simeq L_\text{max}/5$ in case of a Kolmogorov power spectrum, assuming $L_{\rm max}\gg L_{\rm min}$.

In our simulations, we are not sensitive to the individual parameter values of particle rigidity $\mathcal{R}$, correlation length $\Lc$ (or equivalently outer scale $L_\text{max}$) and rms B-field, but only to the combination $\rg/\Lc$. Therefore, our simulations can be applied to different combinations of these parameters. However, when considering the use for a particular physics case, e.g.\ transport of Galactic CRs, we oftentimes adopt particular fiducial parameter values. Specifically, we consider an outer scale $L_\text{max} = 150 \, \text{pc}$~\citep{2010ApJ...714.1398C} which for Kolmogorov turbulence corresponds to a correlation length of $\Lc \simeq 30 \, \text{pc}$. Smaller values have been inferred for interarm regions~\citep{2008ApJ...680..362H}. With an rms value of $4 \, \mu\text{G}$, the gyroradius evaluates to $\rg = 0.270 \, \text{pc} (\mathcal{R}/\text{PV}) (B_{\text{rms}}/4 \, \mu\text{G})^{-1}$. 

It is important to recognise also that in individual realisations of the turbulent magnetic field, modes with wavelengths much longer than the gyroradius under consideration also create a large-scale, coherent contribution to the total magnetic field. However, in contrast to background field $\vct{B}_0$, the strength and direction of this contribution is not fixed and varies between different realisations of the turbulent magnetic field. We will refer to the sum of the background field $\vct{B}_0$ and these additional contribution as the \textit{effective background field} $\vct{B}_{\text{eff}}$. In Sec.~\ref{sec:paacf} and \ref{sec:running_diff_coeffs}, we will highlight some effects due to the effective background field. 

We note also that some of the previous works on cosmic-ray transport in isotropic turbulence have explored also the case where the spectrum is non-zero below $2\pi/L_{\rm max}$ \citep[see e.g.][]{2007JCAP...06..027D,2016MNRAS.457.3975S} which might result in a slightly larger correlation length for the same turbulence level and correspondingly lead to a slightly larger perpendicular mean free. We expect, however, that the rigidity dependence of the particle mean free paths is qualitatively similar to the one for the spectrum adopted in this work. 

Finally, we note that physical quantities referred to as observables resulting from simulations are oftentimes evaluated by performing an average over initial particle directions (see Sec.~\ref{sec:observables}). In this work, the number of test particles are always between $3\times 10^3$ and $8\times 10^5$ depending on the values of the turbulence levels and reduced rigidities. In fact, for the low rigidity regime that we are most interested in, the number of test particles was chosen to be $8\times 10^5$ for the lowest four rigidities $\rg/\Lc = 8.90\times 10^{-5}, 4.45\times 10^{-4}, 8.90\times 10^{-4}$, and $2.67\times 10^{-3}$.

\subsection{Generating synthetic turbulence \label{sec:synthetic_turbulence}}

In the following, we will review briefly the generation of synthetic, isotropic magnetic field turbulence on a computer, using the so-called harmonic method and nested grid method (see \citet{2020Ap&SS.365..135M} for a more thorough discussion). 

\subsubsection{Harmonic Method}

In the harmonic method, the turbulent magnetic field is computed for each position $\vct{r}$ as the superposition of $N$ plane waves with (random) amplitudes $A_n$, phases $\beta_n$ and polarisations $\uvct{\xi}_n$. The wave vectors are split into a normalization and a random direction $\vct{k}_n = k_n \uvct{k}_n$. The superposition of plane waves can then be written as
\begin{equation}
\label{eq:planewave2}
\vct{\delta B}(\vct{r}) = \sqrt{2}\sum_{n=0}^{N-1} A_n \uvct{\xi}_n \cos{(k_n\uvct{k}_n\cdot\vct{r}+\beta_n)}.
\end{equation}
For isotropic turbulence, the $\uvct{k}_n$ are isotropically distributed, 
\begin{equation}
\uvct{k}_n = 
\begin{pmatrix}
\sqrt{1-\eta_n^2}\cos{\phi_n}\\
\sqrt{1-\eta_n^2}\sin{\phi_n}\\
\eta_n
\end{pmatrix},
\end{equation}
with $\phi_n$ sampled uniformly in $[0,2\pi[$ and $\eta_n$ sampled uniformly in $[-1,1]$. 
Concerning the polarization vector, there exist different parametrizations but we will adopt the one from Tautz and Dosch~\citep{2013PhPl...20b2302T} for some systematic studies in Appendix \ref{sec:harmonic_vs_grid}. The polarization vector they suggested is given by
\begin{equation}
\uvct{\xi}_n= 
\begin{pmatrix}
-\sin{\alpha_n}\sin{\phi_n}+\cos{\alpha_n}\cos{\phi_n}\eta_n\\
\sin{\alpha_n}\cos{\phi_n}+\cos{\alpha_n}\sin{\phi_n}\eta_n\\
-\sqrt{1-\eta_n^2}\cos{\alpha_n}
\end{pmatrix}.
\end{equation}
If $\alpha_n$ were set to $0$, this would result in Alfvénic polarisation; if it were set to $\pi/2$, this would lead to magnetosonic polarisation. Here, we draw $\alpha_n$ from a uniform distribution between $[0,2\pi[$. In the literature this is referred to as ``isotropic turbulence'' and constitutes an equal mixture of Alfvénic and magnetosonic polarisations. 

In total, this method requires saving six real numbers per mode $n$. The number of wave modes typically adopted for a dynamical range of $k_\text{min}/k_\text{max}\sim 10^4$ is $N=\mathcal{O}(100)-\mathcal{O}(1000)$~\citep{2020Ap&SS.365..135M}.

\subsubsection{Nested Grid Method in Isotropic Turbulence\label{sec:nested_grid_method}}

The major drawback of the harmonic method is that while in principle it is very accurate, at each position, for every particle the sum has to be evaluated up to large $N$ which is computationally very expensive. To circumvent this problem grid-based methods can be used~\citep{2002GeoRL..29.1048Q,2007JCAP...06..027D,2016MNRAS.457.3975S}. In the grid method, the magnetic field is generated only once on a large grid that is then repeated periodically to span the simulation volume. For each particle and time step, the magnetic field is then interpolated between the closest grid points.

Since the magnetic field in three-dimensional isotropic turbulence depends on three spatial components the memory requirements of the grid-based method grow as the number of grid points cubed. Simulating a turbulent field with large enough dynamical range for the particle rigidities considered can thus quickly exceed the computer memory. It has therefore been suggested~\citep{2012JCAP...07..031G} to use a number of nested grids, each grid corresponding to a certain range in wave number. The magnetic field can be calculated as the superposition of the different magnetic field grids in position space where the grids are repeated periodically over the entire simulation volume. The nested grid method is also the technique that we adopted for generating synthetic turbulence and is at the basis of all the main numerical results presented Sec. \ref{sec:results}.

While evaluating the magnetic field in the grid-based method is computationally much less expensive, it requires a prohibitively large amount of memory to cover a large dynamical range. This memory requirement depends strongly on the dynamical range that is needed for the problem. We will assume a particle rigidity of $10 \, \text{TV}$ and an outer scale of turbulence of $L_{\text{max}} = 150 \, \text{pc}$. The smallest wavelength on the grid has to be at least as small as the gyroradius at this energy. Assuming a field strength of $B_{\text{rms}} = 4 \, \mu\text{G}$ this means $r_g/L_{\text{max}} = 1.78 \times 10^{-5}$. To avoid artefacts due to periodicity, the grid should be larger than $L_{\text{max}}$ by a factor of a few, let's say $4$. Also, it has proven necessary in practice to have a smallest wavelength that is at least a factor 10 smaller than the gyroradius of the particle.

This means that the number of grid points required to span this dynamical range is at least $N = 10 \times 4 \times 2 / (1.78 \times 10^{-5}) \simeq 4,494,382$. Accounting for the requirements of the FFT algorithms~\citep{Cooley:1965zz,Johnson_Frigo_2008} that the number of grid points should be a power of $2$ this increases $N$ to $N = 2^{23} = 8,388,608$ per dimension. For isotropic turbulence this results in a total $N^3 = 590,295,810,358,705,651,712$ grid points for each field direction. The total memory requirement, assuming $8 \, \text{B}$ per double precision floating point number, then evaluates to $14,167,100 \, \text{PB}$ which makes such simulations currently unfeasible.

To get around this constraint \citet{2012JCAP...07..031G} proposed to superimpose multiple grids with different sizes and therefore different resolutions. The small wavelengths are resolved on the smallest grid with the finest resolution while the modes with large wavelengths are resolved on a larger grid with larger grid spacing.

To implement this method of generating the turbulent magnetic field the dynamical range $[k_{\text{min}}, k_{\text{max}}]$ is split into $G$ different grids that are set up as described previously with each grid having a dynamical range $[k_g,k_{g+1}]$, where $k_0 = k_{\text{min}} = 2 \pi / L_{\text{max}}$ and $k_G = k_{\text{max}} = 2\pi / L_{\text{min}}$. In addition to the modes that contribute to the power spectrum, each grid also includes some modes without power as padding. This means that the grids are both larger and have a finer resolution than the theoretical minimum requirement to resolve the modes. This is necessary to avoid artefacts due to discreteness effects and grid periodicity.

It is well known that the largest wavenumber that can theoretically be resolved on a grid with finite resolution  is given by the Nyquist wavenumber $k_{\text{Nyquist}} =  (1/2) 2 \pi / \Delta x$~\citep{1928TAIEE..47..617N}. Waves with wavenumbers larger than this will, due to aliasing effects, look like waves with smaller wavenumbers traveling in the opposite direction~\citep{2016era..book.....C}. Additionally, a factor of a few is included to ensure that the modes with the smallest wavelengths are always resolved by enough grid points. To avoid grid periodicity effects one should also make sure that the largest wavelength fits on the grid at least a few times.

For example, grids at each level $g$ have sizes $L_g$ and $N=128$ grid points, implying $\Delta x_g = L_g / N$. However, we choose for only the modes with wavelengths between $L_{g,\text{min}} = L_g / 20$ and $L_{g,\text{max}} = L_g / 8$ to have non-vanishing amplitude. The next grid level ($g+1$) is defined such that $L_{g+1,\text{max}} = L_{g,\text{min}}$. Specifically, for the use case with $\rg/\Lc = 8.91 \times 10^{-5}$, we have chosen 16 grid levels. At level $g=0$, the grid has a size $L_0 = 40 \, \Lc$, $L_{0,\text{max}} = 5 \, \Lc$ and $L_{0,\text{min}} = 2 \, \Lc$. At level $g=1$, $L_1 = 16 \, \Lc$, $L_{1,\text{max}} = 2 \, \Lc$ and $L_{1,\text{min}} = 0.8 \, \Lc$ and so on. Finally, at level $g=15$, $L_{15} = 4.3 \times 10^{-5} \, \Lc$, $L_{15,\text{max}} = 5.4 \times 10^{-6} \, \Lc$ and $L_{15,\text{min}} = 2.1 \times 10^{-6} \, \Lc$. The dynamic range $L_{0,\text{max}} / L_{15,\text{min}}$ covered is thus more than six orders of magnitude. 

To ensure a continuous power spectrum, the amplitudes on each grid $g$ have to be rescaled such that they contribute the correct fraction of the total power $\delta B^2$ to the power spectrum. It follows that for a pure power law power spectrum the variance $\delta B_g^2$ of an individual grid containing modes in $[L_{g,\text{min}}, L_{g,\text{max}}]$ needs to be rescaled,
\begin{equation}
\delta B_g^2 \propto \delta B^2 \frac{L_{g,\text{min}}^{\gamma-1} - L_{g,\text{max}}^{\gamma-1}}{L_{\text{max}}^{\gamma-1} - L_{\text{min}}^{\gamma-1}} \, .
\end{equation}

Note that in particular for small reduced rigidities, at the end of our simulations, only a very small fraction of particles will have travelled further than a few times $L_0$. 
We therefore consider artefacts due to the grid periodicity to be negligible.

\subsubsection{ODE solver\label{sec:solvers}}

To calculate the test particle trajectories the Newton-Lorentz equation~\eqref{eqn:Newton-Lorentz} with a vanishing electric field has to be solved. Formally, this is a second order differential equation in time. To solve this numerically it can be rewritten as a set of two first order differential equations,
\begin{align}
\frac{\dd \vct{r}}{\dd t} &= \vct{v} \, , \\
\frac{1}{\Omega} \frac{\dd \uvct{v}}{\dd t} &= \uvct{v} \times \uvct{B} \, .
\end{align}
Here, we have used that $\gamma=\text{const.}$ and introduced the relativistic gyrofrequency $\Omega \equiv q B / (\gamma m c)$.

The generic problem of solving a first order ordinary differential equation is a well studied problem~\citep{PresTeukVettFlan92}, with a multitude of differential equation solvers available. The most widely used solver for ordinary differential equations is probably the Runge-Kutta solver~\citep{runge1895,runge1901} that is already implemented for example in the \texttt{odeint} package of the \texttt{boost} library~\citep{2011AIPC.1389.1586A}. It should be noted, however, that in this method errors on the particle velocity can accumulate and over a large number of steps the particle energy can change significantly.

The Hamiltonian of a charged particle in a stationary magnetic field is given by~\citep{1995iqm..book.....G}
\begin{equation}
H(\vct{r}, \vct{p}) = \frac{1}{2} \left( \vct{p} - q \vct{A}(\vct{r}) \right)^2 \, ,
\end{equation}
with the vector potential $\vct{A}(\vct{r})$ defined by $\vct{B}(\vct{r}) = \vct{\nabla} \times \vct{A}(\vct{r})$. It is easy to see that since this Hamiltonian does not explicitly depend on time, the total energy of the particles should be conserved. While there are symplectic differential equation solvers that inherently conserve particle energy, they typically require that the Hamiltonian is separable $H(\vct{r}, \vct{p}) = T(\vct{p}) + V(\vct{r})$. Due to its ease of implementation and inherent ability to conserve energy, the differential equation solver that has become the standard in charged particle propagation is the Boris solver~\citep{boris,BirdsallLangdon}. It has been pointed out that, even though this algorithm is technically not symplectic, it conserves phase-space volume, which is a necessary but not a sufficient condition for symplectic algorithms~\citep{2013PhPl...20h4503Q}. It can be shown that the error on the energy has a global upper bound for all time steps~\citep{2018APS..DPPTO8003H}. For the Runge-Kutta method, we have used the Runge-Kutta-Dopri5 solver of the \texttt{odeint} package in the \texttt{boost} library~\citep{2011AIPC.1389.1586A}. A comparison of different differential equation solvers for this problem can be found in \citet{2018ApJS..235...21R}. We also compare results for the Boris-Push and for the Runge-Kutta solver in Appendix~\ref{sec:test_solvers}. 

In order to study the transport of magnetic field lines, we have also solved the equations for a field line $\vct{r}_{\text{FL}}(s) = (x(s), y(s), z(s))^T$, 
\begin{align}
\frac{\dd x}{\dd s} &= \frac{\delta B_x}{\sqrt{B_0^2 + \delta B^2}} \, , \\
\frac{\dd y}{\dd s} &= \frac{\delta B_y}{\sqrt{B_0^2 + \delta B^2}} \, , \\
\frac{\dd z}{\dd s} &= \frac{B_0 + \delta B_z}{\sqrt{B_0^2 + \delta B^2}} \, . 
\end{align}
For the field-line integration, we have employed the Runge-Kutta-Dopri5 solver.

\subsection{Observables}
\label{sec:observables}

\subsubsection{Pitch-angle auto-correlation function}

Particle motion becomes diffusive as soon as individual particle velocities become uncorrelated with their initial velocities. For diffusion along the background magnetic field, the relevant velocity component is the parallel one, $v_{\parallel} = v \mu$, $\mu$ being the cosine of the pitch-angle. The decorrelation of the pitch-angle (cosine) due to interactions with the turbulent magnetic field is known as pitch-angle scattering. 

The decorrelation of the parallel velocity component can be investigated by considering the pitch-angle autocorrelation function $C(t)$~\citep{2001PhRvD..65b3002C},
\begin{equation}
C(t) \equiv \frac{\langle \mu_0 \mu(t) \rangle}{\langle \mu_0^2 \rangle} \, ,
\label{eqn:def_paac}
\end{equation}
where $\mu_0 \equiv \mu(0)$. The angled brackets $\langle \mathellipsis \rangle$ here denote averages over initial particle directions and over the ensemble of turbulent magnetic fields. From this the scattering time can be defined as the coherence time of the pitch-angle cosine $\mu$,
\begin{equation}
\taus \equiv \int_0^{\infty} \mathrm{d}\tau \, C(\tau) \, . \label{eqn:scatteringtime}
\end{equation}

It has previously been argued~\citep{2001PhRvD..65b3002C} that the pitch-angle autocorrelation function $C(t)$ takes an exponential form, $C(t) = \exp[ - t / \tau]$. For this form, the time constant $\tau$ is immediately identified as the scattering time. In the following, we will generalise this to $C(t) = A \exp[ - t / \tau]$ with $A \leq 1$. For $A < 1$, the scattering time $\taus$ will be suppressed with respect to $\tau$, that is $\taus = A \tau < \tau$.

\subsubsection{Particle diffusion coefficients}

The nature of particle transport, in particular the nature of transport of charged CRs in turbulent magnetic fields, can be characterised by the time-dependence of the mean-square displacement in different directions $i$, $\langle ( \Delta r_i )^2 \rangle$. If this is of power law form, that is
\begin{equation}
\langle (\Delta r_i)^2 \rangle = \left\langle \left( r_i(t) - r_i(0) \right)^2 \right\rangle \propto t^\alpha \, ,
\label{eqn:diffclass}
\end{equation}
transport is called sub-diffusive if $0<\alpha<1$, diffusive for $\alpha = 1$, super-diffusive for $1<\alpha<2$ and ballistic for $\alpha = 2$~\citep{2009ASSL..362.....S,2020Ap&SS.365..135M}. The diffusion coefficient $\kappa$, that is the constant of proportionality in Eq.~\eqref{eqn:diffclass} for the case of diffusive transport, plays a central role in the transport theory of charged CRs. In this section, we will describe various methods for how it can be derived, for instance from test particle simulations.

We define the running diffusion coefficient with respect to direction $i$ as
\begin{equation}
d_{ii}(t) \equiv \frac{1}{2} \frac{\dd}{\dd t} \langle (r_i(t) - r_i(0))^2 \rangle \, .
\label{eqn:def_dii}
\end{equation}
Specifically, since the isotropy of the space is broken by the presence of a background magnetic field, here assumed to point in the $z$-direction, we distinguish the parallel and the perpendicular running diffusion coefficients,
\begin{align}
d_{\parallel}(t) &\equiv \frac{1}{2} \frac{\dd}{\dd t} \langle (z(t) - z(0))^2 \rangle \, , \label{eqn:def_dpar} \\
d_{\perp}(t) &\equiv \frac{1}{4} \frac{\dd}{\dd t} \left( \langle (x(t) - x(0))^2 \rangle + \langle (y(t) - y(0))^2 \rangle \right) \, .
\label{eqn:def_dperp}
\end{align}
If transport is diffusive at late times, those converge towards the asymptotic diffusion coefficients
\begin{align}
\kappa_{\parallel} \equiv \lim_{t \to \infty} d_{\parallel}(t) \, , \label{eqn:def_kappa_par} \\
\kappa_{\perp} \equiv \lim_{t \to \infty} d_{\perp}(t) \, .\label{eqn:def_kappa_perp}
\end{align}

The TGK formalism~\citep{taylor1922,1954JChPh..22..398G,1957JPSJ...12..570K} allows to relate $\kappa_{\parallel}$ and $\kappa_{\perp}$ to the velocity autocorrelation functions. In particular,
\begin{equation}
\langle (\Delta r_i)^2 \rangle = 2 \int_0^t \dd t' \, (t-t') \langle v_i(0) v_i(t') \rangle \, ,
\end{equation}
and for the asymptotic diffusion coefficients
\begin{align}
\kappa_{\parallel} &= \int_0^{\infty} \dd t' \, \langle v_z(0) v_z(t') \rangle \, , \\
\kappa_{\perp} &= \int_0^{\infty} \dd t' \, \langle v_x(0) v_x(t') \rangle = \int_0^{\infty} \dd t' \, \langle v_y(0) v_y(t') \rangle .
\end{align}

For an isotropic particle distribution $\langle \mu^2 \rangle = 1/3$ and so we can relate the scattering time $\taus$, eq.~\eqref{eqn:scatteringtime}, and the parallel diffusion coefficient $\kappa_{\parallel}$ via the pitch-angle autocorrelation function $C(t) = 3 \langle \mu(t) \mu_0 \rangle = 3/v^2 \langle v_z(0) v_z(t) \rangle$,
\begin{equation}
\taus = \!\! \int_0^{\infty} \mathrm{d}\tau \, C(\tau) = \frac{3}{v^2} \! \int_0^{\infty} \dd t' \langle v_z(0) v_z(t') \rangle = \frac{3 \kappa_{\parallel}}{v^2} . \label{eqn:def_taus}
\end{equation}
This coincides with the time at which a ballistic running diffusion coefficient, $d_{\parallel}(t) = (v^2/3) t$ intersects with the asymptotic diffusion coefficient $\kappa_{\parallel}$. 

Before concluding this section, we note that an alternative definition of the running diffusion coefficients is sometimes employed in the literature,
\begin{align}
d_{\parallel,\text{frac}}(t) \! &\equiv \! \frac{1}{2t} \langle (z(t) - z(0))^2 \rangle \, , \label{eqn:def_dpar_frac} \\
d_{\perp,\text{frac}}(t) \! &\equiv \! \frac{1}{4t} \left( \langle (x(t) - x(0))^2 \rangle \! + \! \langle (y(t) - y(0))^2 \rangle \right) \! . \label{eqn:def_dperp_frac}
\end{align}
While the derivative definition of eqs.~\eqref{eqn:def_dpar} and \eqref{eqn:def_dperp} agrees with the fractional definition of eqs.~\eqref{eqn:def_dpar_frac} and \eqref{eqn:def_dperp_frac} for the diffusive limit, in general the running diffusion coefficients differ. In particular, the fractional definition tends to smoothen features. In the following, we will adopt the derivative definition although this will be more susceptible to noise, but we discuss also possible differences in Appendix~\ref{sec:deriv_frac}.

\subsubsection{Field line diffusion coefficients\label{sec:FL_diffusion_coeff}}

In the same way that we have characterised particle transport through the power law dependence on time $t$ of the mean-square displacement, we can also characterise the transport of magnetic field lines through the power law dependence of the mean-square displacement of the field line $\vct{r}(s)$ on arc-length $s$, 
\begin{equation}
\langle (\Delta r_i)^2 \rangle = \left\langle \left( r_i(s) - r_i(0) \right)^2 \right\rangle \propto s^\alpha \, .
\label{eqn:FLclass}
\end{equation}
Specifically, we say that field line transport is sub-diffusive if $0<\alpha<1$, diffusive for $\alpha = 1$, super-diffusive for $1<\alpha<2$ and ballistic for $\alpha = 2$. 

The running (perpendicular) field line diffusion coefficient with respect to $s$ is defined as 
\begin{equation}
d_{\text{FL}}(s) \! \equiv \! \frac{1}{4} \frac{\dd}{\dd s} \left( \langle (x(s) - x(0))^2 \rangle \! + \! \langle (y(s) - y(0))^2 \rangle \right) \! . \label{eqn:def_dFL_s}
\end{equation}
with the asymptotic value
\begin{equation}
\kappa_{\text{FL}} \equiv \lim_{s \to \infty} d_{\text{FL}}(s) \, .
\end{equation}
Analogously, a parallel field line diffusion coefficient could be defined, but we will not be needing this.

Note that oftentimes low turbulence levels are assumed such that the field lines are almost parallel to the $z$-direction and $s \approx z$. In that case, the running (perpendicular) field line diffusion coefficient with respect to $z$ is defined as 
\begin{equation}
d'_{\text{FL}}(z) \! \equiv \! \frac{1}{4} \frac{\dd}{\dd z} \! \left( \langle (x(z) - x(0))^2 \rangle \! + \! \langle (y(z) - y(0))^2 \rangle \right) \! . \label{eqn:def_dFL}
\end{equation}
with the asymptotic value
\begin{equation}
\kappa'_{\text{FL}} \equiv \lim_{z \to \infty} d'_{\text{FL}}(z)  \, .
\end{equation}

Given that we extract field line positions $\vct{r}_{\text{FL}}(s) = (x(s), y(s), z(s))^T$ for fixed values of $s$ from the simulations, we have computed the vertical field line displacements as functions of $z$, that is $x(z)$ and $y(z)$, by computing the mean $x$ and $y$ for given intervals in $z^2$. Specifically, we consider $51$ bins, logarithmically spaced between $z^2 = 1.6 \times 10^{-6} \Lc^2$ and $1.6 \times 10^{6} \Lc^2$. While this introduces a certain degree of smoothing into the data, we judge the loss of precision to be acceptable.

\section{Results\label{sec:results}}
 
\subsection{Pitch-angle auto-correlation\label{sec:paacf}}

\begin{figure*}
\includegraphics[width=\textwidth]{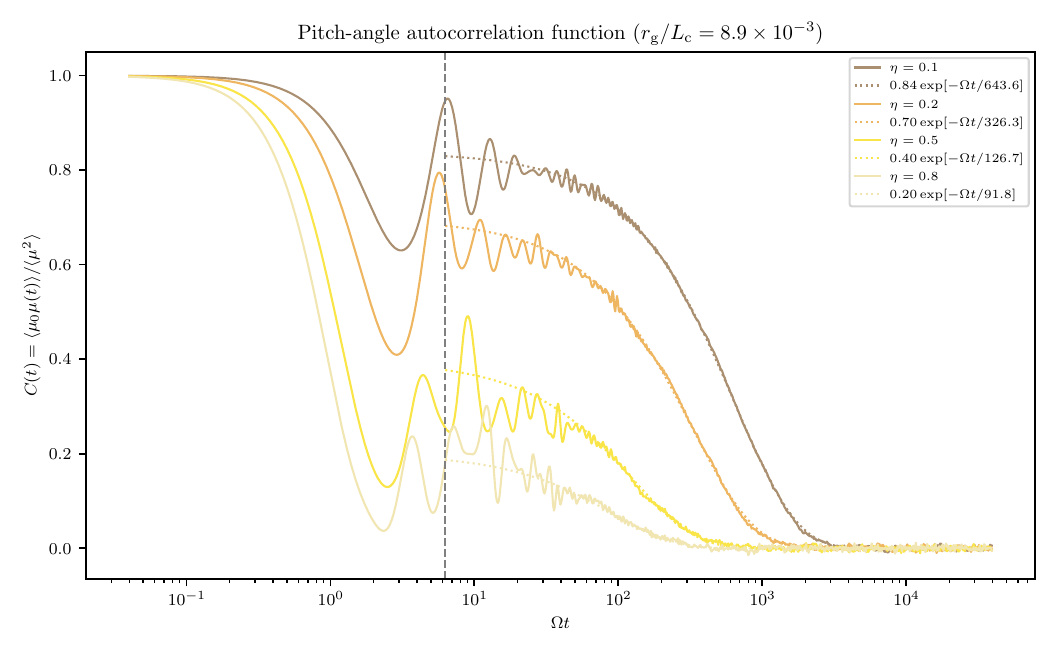}
\caption{Pitch-angle autocorrelation function $C(t)$ for $\rg / \Lc = 8.9 \times 10^{-3}$ and different turbulence levels $\eta$. The dashed vertical line indicates $\Omega t = 2 \pi$. The dotted lines are fits of the form $A \exp[-t / \tau]$ to the different $C(t)$ for $\Omega t > 2 \pi$.}
\label{fig:paac_1PeV_different_eta}
\includegraphics[width=\textwidth]{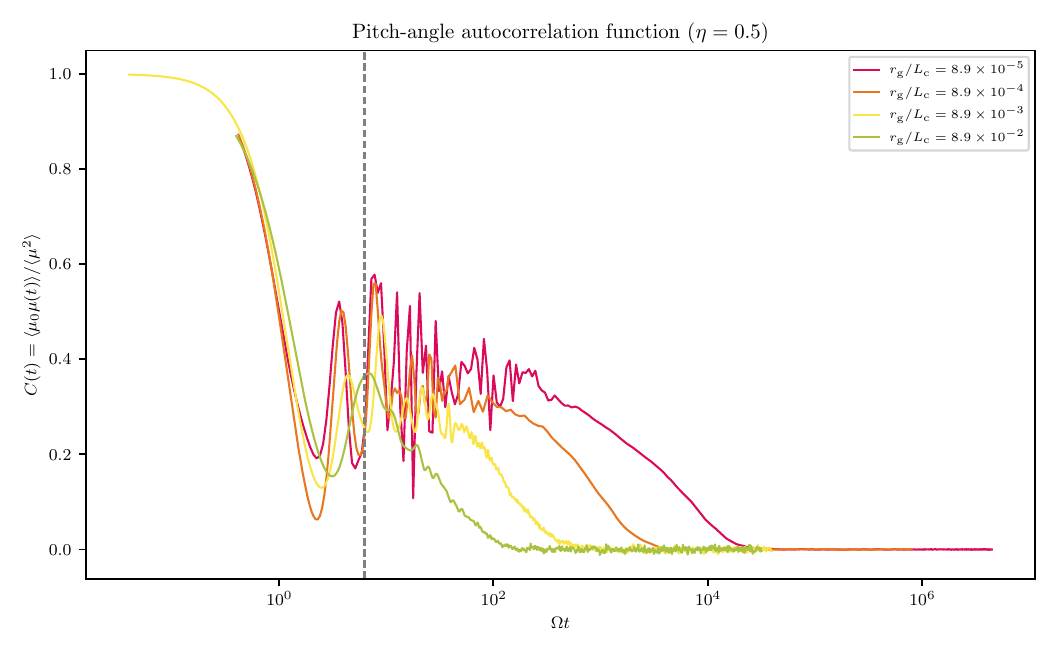}
\caption{Pitch-angle autocorrelation function $C(t)$ for $\eta = 0.5$ and different reduced rigidities $\rg / \Lc$. The dashed vertical line indicates $\Omega t = 2 \pi$.}
\label{fig:paac_different_rigidities_eta_0p5_loglin}
\end{figure*}

We start the discussion of our results by looking at the pitch-angle autocorrelation $C(t)$, see eq.~\eqref{eqn:def_paac}. In Fig.~\ref{fig:paac_1PeV_different_eta} we show $C(t)$ as a function of $t$ for different turbulence levels $\eta$, see eq.~\eqref{eqn:def_eta} at the fixed reduced rigidity $\rg{}/\Lc = 8.9 \times 10^{-3}$ which for our fiducial parameters ($\Lmax = 150 \, \text{pc}$, $\Brms = 4 \, \mu\text{G}$, see Sec.~\ref{sec:turbulence}) corresponds to a rigidity of $1 \, \text{PeV}$. We have chosen this rigidity since for $\rg{}/\Lc = 8.9 \times 10^{-3}$, most particles can resonantly interact with the turbulent magnetic field. 

Three separate phases can be distinguished in the time-dependence of the pitch-angle autocorrelation.

\subsubsection{Early times ($\Omega t \ll 1$)}

For $\Omega t \ll 1$, the particles are moving ballistically into their initial direction and are not yet affected by the magnetic field. The correlation of all velocity components, in particular of the $z$-component $\mu = v_z / | \vct{v}|$ is therefore one,
\begin{equation}
C(t) = \frac{\langle \mu_0 \mu(t) \rangle}{\langle (\mu_0)^2 \rangle} \simeq \frac{\langle \mu_0 \mu_0 \rangle}{\langle (\mu_0)^2 \rangle} = 1 \, .
\label{eqn:paac_early}
\end{equation}

\subsubsection{Intermediate times ($1 \ll \Omega t \ll \Omega \taus$)}

For $1 \ll \Omega t \ll \Omega \taus$, particles have started to gyrate in the effective background magnetic field. This results in a plateau in the pitch-angle autocorrelation function: For low turbulence levels, the suppression is moderate, for large turbulence levels it is much stronger. Depending on the initial pitch-angle and the angle $\thetaeff$ between $\vct{B}_{\text{eff}}$ and the $z$-direction, the correlation between $\mu_0$ and $\mu(t)$ oscillates between a suppressed value and $1$. Imagine, for instance, the time-dependence of the $z$-component of the particle velocity, $v_z(t)$, for the extreme case that the effective background magnetic field is pointing along the $x$-direction. (This could be the case for a large turbulence level where $B_0$ is subdominant.) In that case, the $z$-component would oscillate between its initial value $v_z(0)$ and the negative opposite, $-v_z(0)$. In the average over all initial particle directions and over the ensemble of turbulent magnetic fields, this would result in a suppression of the pitch-angle autocorrelation $C(t)$.

We can formalise this reasoning as follows. Consider the initial (normalised) velocity $\hat{\vct{v}}_0 = (\sqrt{1 - \mu_0^2} \cos \phi_0, \sqrt{1 - \mu_0^2} \sin \phi_0, \mu_0)^T$, parametrised by the initial pitch-angle $\mu_0$ and azimuth $\phi_0$. Without loss of generality, we take the effective background field to point in the direction $(\sin \thetaeff, 0, \cos \thetaeff)^T$. The gyration around this direction by an angle $\psi = \Omega t$ results in the pitch-angle cosine 
\begin{equation}
\begin{aligned}
\mu(t) = &\sqrt{1 - \mu_0^2} \sin \thetaeff \sin \phi_0 \sin \Omega t \\
&-\sqrt{1 - \mu_0^2} \sin \thetaeff \cos \thetaeff \cos \phi_0 (\cos \Omega t - 1) \\
&+ \mu_0 (\cos^2 \thetaeff + \cos \Omega t \sin^2 \thetaeff) \, .
\end{aligned}
\end{equation}
The pitch-angle correlation then computes as (assuming that particles are initially launched isotropically) 
\begin{align}
C(t) &= 3 \left\langle \mu_0 \mu(t) \right\rangle_{0} \\
&= 3 \frac{1}{4 \pi} \int_0^{2 \pi} \dd \phi_0 \, \int_{-1}^1 \dd \mu_0 \, \mu_0 \mu(t) \\
&= \cos^2 \thetaeff + \sin^2 \thetaeff \cos \Omega t \, ,
\label{eqn:paac_intermediate}
\end{align}
where the angled brackets $\langle \cdot \rangle_0$ denote the average over the initial directions.

For the distribution of effective background field directions, we adopt a von Mises-Fisher distribution~\citep{1953RSPSA.217..295F},
\begin{equation}
F(\thetaeff) \equiv \sigmaeff^{-2} \csch \left( \sigmaeff^{-2} \right) \exp \left[ \sigmaeff^{-2} \cos\thetaeff \right] / 2 \, ,
\end{equation}
an extension of a 2D Gaussian to the sphere. We estimate the dispersion $\sigmaeff$ of the $\thetaeff$-distribution to be 
\begin{equation}
\tan^2 \sigmaeff = \frac{\delta B^2}{B_0^2} \, .
\end{equation}
The ensemble-averaged pitch-angle autocorrelation is then given by the Fisher-distribution-weighted average of eq.~\eqref{eqn:paac_intermediate}, 
\begin{align}
\langle C(t) \rangle_{\thetaeff} &\equiv \int_0^{\pi} \dd \thetaeff \sin \thetaeff F(\thetaeff) C(t) \\
&= 1 - 2 \sigmaeff^4 \cos (\Omega t) + 2 \sigmaeff^4 \\ 
& + 2 \sigmaeff^2 \coth \left(\frac{1}{\sigmaeff^2}\right) (\cos (\Omega t)-1) \, .
\end{align}
Finally, averaging over gyroperiods,
\begin{equation}
\langle C(t) \rangle_{\thetaeff, t} = 1 + 2 \sigmaeff^4 - 2 \sigmaeff^2 \coth \left(\frac{1}{\sigmaeff^2}\right) \, .
\end{equation}
For the turbulence levels adopted in Fig.~\ref{fig:paac_1PeV_different_eta}, that is $\eta=0.1, 0.2, 0.5$ and $0.8$, we find $A = \langle C(t) \rangle_{\thetaeff, t} = 0.81, 0.66, 0.43$ and $0.36$, respectively. 

\subsubsection{Late times ($\Omega t \gg \Omega \taus$)}

Finally, for $\Omega t \gg \Omega \taus$, the pitch-angle autocorrelation is exponentially suppressed due to pitch-angle scattering. For low turbulence levels, the time constant, that is the scattering time $\taus$ is large, for large turbulence levels, it becomes smaller, as expected. 

\subsubsection{Comparison with simulation results}

We have tested our understanding of the plateau for $\Omega t \gg 1$ by fitting an exponential function $A \exp[ -t / \taus]$ to the pitch-angle autocorrelation function $C(t)$. These fits are indicated by the dashed lines in Fig.~\ref{fig:paac_1PeV_different_eta}. The parameter values $A$ for $\eta=0.1, 0.2, 0.5$ and $0.8$ are $0.84, 0.70, 0.40$ and $0.20$ which compares favourably with the above predictions. 

In Fig.~\ref{fig:paac_different_rigidities_eta_0p5_loglin}, we also compare the pitch-angle auto correlation functions for the fixed turbulence level $\eta = 0.5$ and different reduced rigidities. We note that the suppression for $1 \ll \Omega t \ll \Omega \taus$ does not depend on the reduced rigidity, as predicted by our reasoning above, as long as $\rg / \Lc \ll 1$. The scattering time, $\taus$, however grows with reduced rigidity, as expected.

\subsection{Running diffusion coefficients\label{sec:running_diff_coeffs}}

\subsubsection{Running parallel diffusion coefficient}

\begin{figure*}[p]
\includegraphics[scale=1]{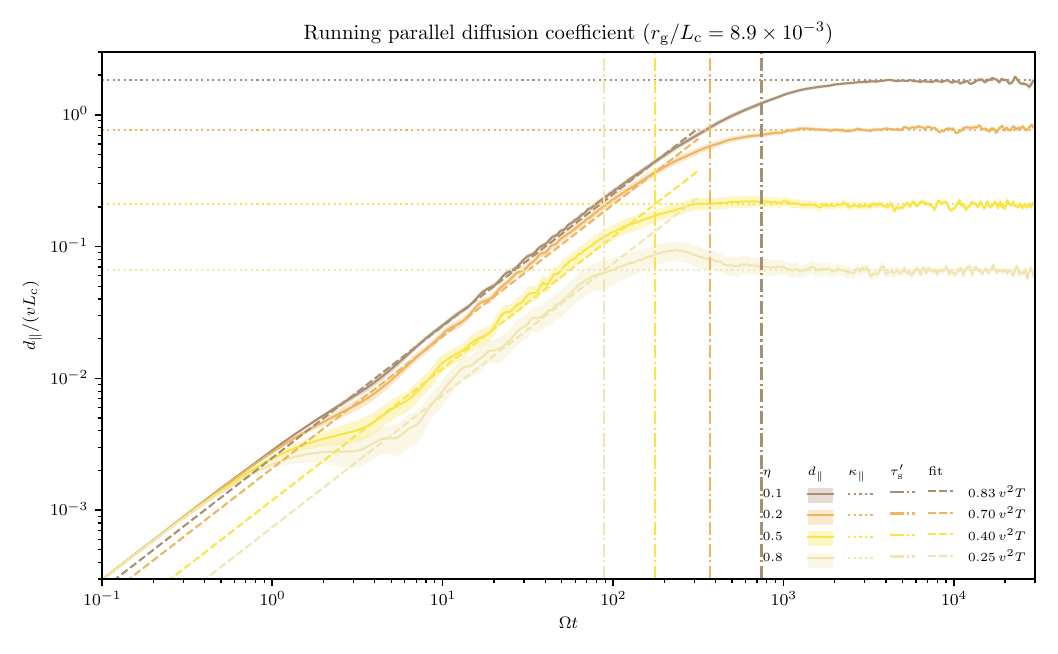}
\caption{
Running parallel diffusion coefficient $d_{\parallel}(t)$ at $\rg/\Lc = 8.9 \times 10^{-3}$ and for various turbulence levels $\eta$. The solid lines show the mean over the ensemble of turbulent magnetic fields, the shaded band indicate the standard mean error. The dashed lines indicate the suppressed ballistic growth. The dotted horizontal lines show the asymptotic diffusion coefficients $\kappa_{\parallel}$ and the vertical dot-dashed lines show the scattering time $\taus'$, increased due to the suppression of the ballistic growth.
}%
\label{fig:dpar_eta}
\includegraphics[scale=1]{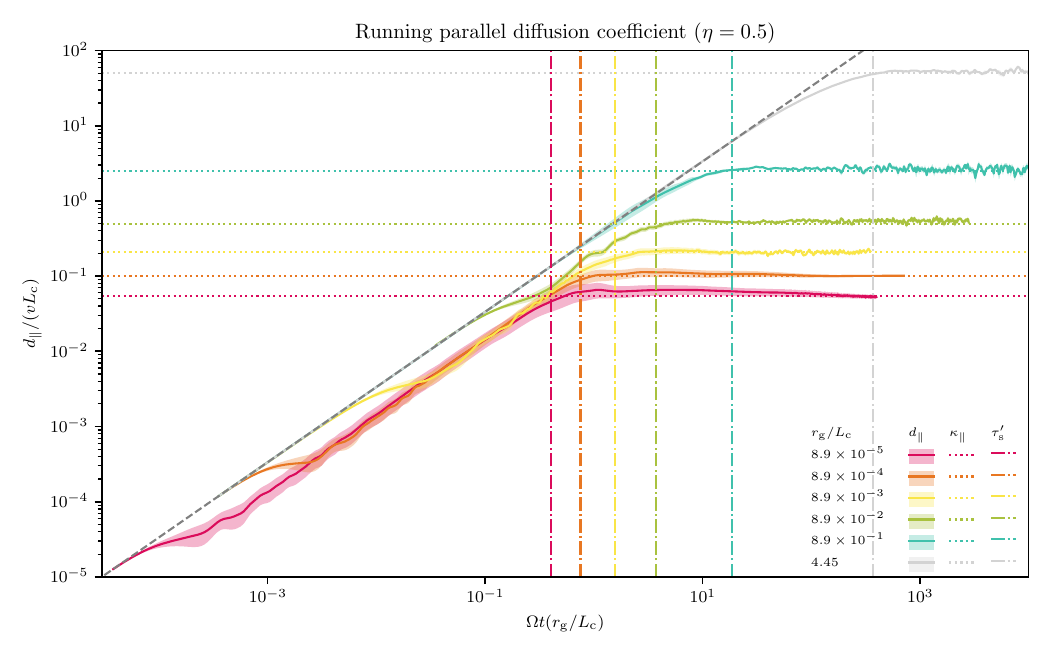}
\caption{
Same as Fig.~\ref{fig:dpar_eta}, but for fixed $\eta = 0.5$ and various reduced rigidities $\rg/\Lc$. The grey dashed line indicates the initial, unsuppressed ballistic phase. 
}%
\label{fig:dpar_rigidities}
\end{figure*}

For the same fixed reduced rigidity as before, $\rg / \Lc = 8.9 \times 10^{-3}$, we illustrate the dependence of the running parallel diffusion coefficient $d_{\parallel}(t)$ on the turbulence level $\eta$ in Fig.~\ref{fig:dpar_eta}. One can identify the same three stages as before and the $\eta$-dependence of the suppression of the pitch-angle auto-correlation function at intermediate times ($1 \ll \Omega t \ll \Omega \taus$) discussed in Sec.~\ref{sec:paacf}. 
\begin{enumerate}
\item For $\Omega t \ll 1$, the particles are moving ballistically into their initial direction and are not yet affected by the magnetic field. The mean square displacement, e.g.\ in the $z$-direction is $\langle (z - z(0))^2 \rangle = (1/3) v^2 t^2$. Therefore, $d_{\parallel}(t) = v^2 t/3$ or $d_{\parallel} /(v \Lc) = (1/3) \Omega t \, \rg/\Lc$. 
\item For $1 \ll \Omega t \ll \Omega \taus$, particles have started to gyrate in the effective background magnetic field. While the mean-square displacement still grows $\propto t^2$, it does so with the rate reduced by the same factor as for the pitch-angle auto-correlation function; for instance, for $\eta = 0.5$, this factor is $\approx 0.4$. 
\item Finally, for $\Omega t \gg \Omega \taus$, pitch-angle scattering is reducing the mean square displacement to diffusive behaviour, $\langle (z - z(0))^2 \rangle \propto t$. Consequently, the running diffusion coefficient, $d_{\parallel}(t)$ flattens out. The asymptotic value $\kappa_{\parallel}(t)$ and the transition time, of course, depend on the reduced rigidity $\rg / \Lc$. We will investigate this rigidity-dependence in more detail in Sec.~\ref{sec:rigidity-dependence}. 
\end{enumerate}
With decreasing turbulence level $\eta$, the scattering time increases and so does the asymptotic diffusion coefficient. 

In Fig.~\ref{fig:dpar_rigidities} we show the running parallel diffusion coefficient $d_{\parallel}(t)$, see eq.~\eqref{eqn:def_dpar}, for different reduced rigidities $\rg / \Lc$ and fixed turbulence level $\eta = 0.5$. Note that we have rescaled the time $\Omega t$ with the reduced rigidities, in order to have all running diffusion coefficients overlap during the ballistic phase despite the different $\Omega$.

\begin{figure*}[p]
\includegraphics[scale=1]{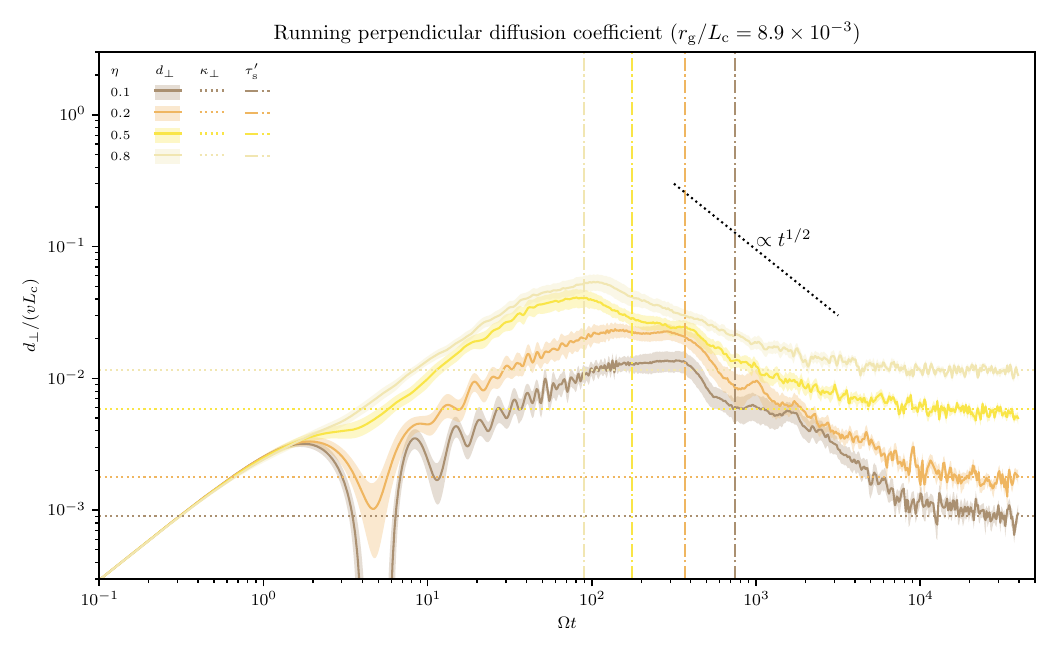}
\caption{Running perpendicular diffusion coefficient $d_{\perp}(t)$ at $\rg/\Lc = 8.9 \times 10^{-3}$ and for various turbulence levels $\eta$. The solid lines show the mean over the ensemble of turbulent magnetic fields, the shaded band indicate the standard mean error. The dashed lines indicate the suppressed ballistic growth. The dotted horizontal lines show the asymptotic diffusion coefficients $\kappa_{\perp}$ and the vertical dot-dashed lines show the scattering time $\taus'$, increased due to the suppression of the ballistic growth.
}%
\label{fig:dperp_eta}
\includegraphics[scale=1]{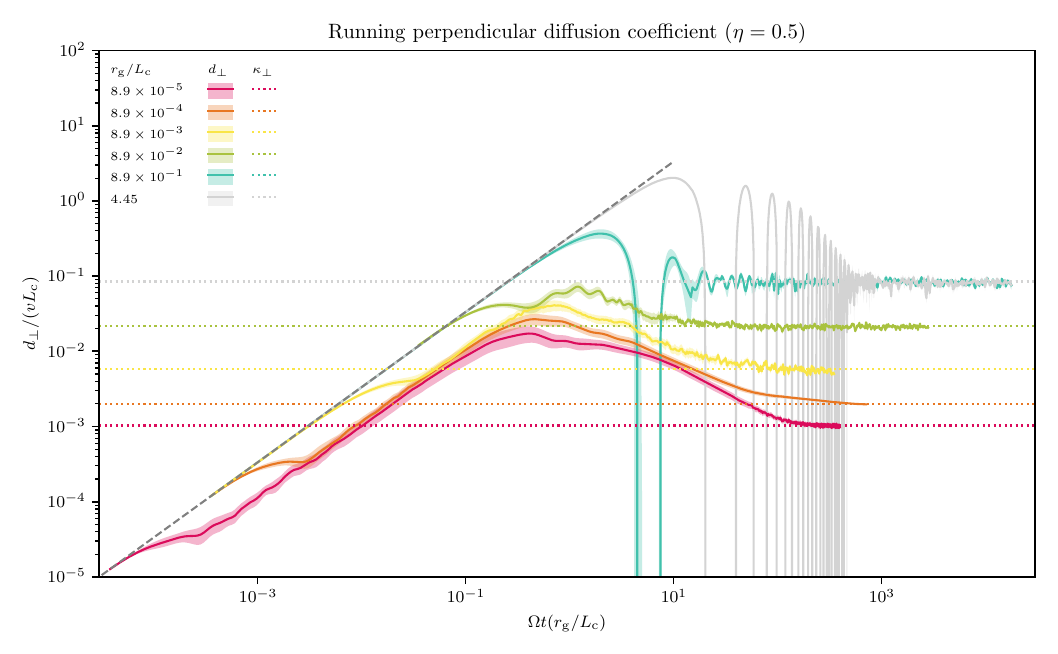}
\caption{
Same as Fig.~\ref{fig:dperp_eta}, but for fixed $\eta = 0.5$ and various reduced rigidities $\rg/\Lc$. The grey dashed line indicates the initial, unsuppressed ballistic phase. 
}%
\label{fig:dperp_rigidities}
\end{figure*}

\begin{figure*}[tbh]
\includegraphics[width=\textwidth]{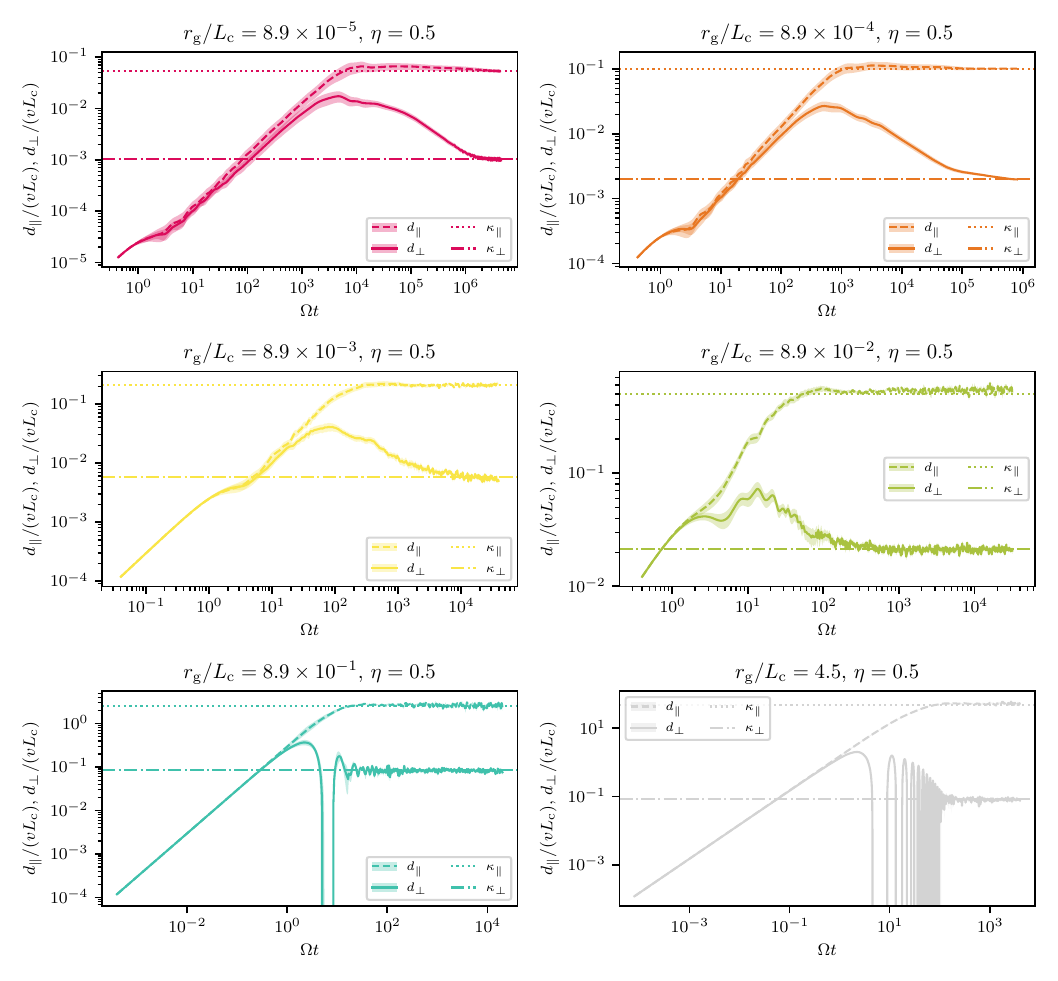}
\caption{Running parallel and perpendicular diffusion coefficients $d_{\parallel}(t)$ and $d_{\perp}(t)$, respectively, for $\eta = 0.5$ and for various reduced rigidities $\rg/\Lc$. The dotted and dash-dotted lines indicate the asymptotic diffusion coefficients $\kappa_{\parallel}$ and $\kappa_{\perp}$, respectively.}
\label{fig:dpar_dperp_rigidities}
\end{figure*}

\subsubsection{Running perpendicular diffusion coefficient}

In Fig.~\ref{fig:dperp_eta} we show the running perpendicular diffusion coefficient $d_{\perp}(t)$, see eq.~\eqref{eqn:def_dperp}, at fixed reduced rigidities $\rg / \Lc=8.9 \times 10^{-3}$ and for different turbulence level $\eta$. The time-dependence is more complicated than that of the pitch-angle auto-correlation function or of the running parallel diffusion coefficient in that we can distinguish four separate phases which are separated by the times $\Omega t = 1,\, \Omega \taus$ and $\Omega \tauc$:
\begin{enumerate}
\item For $\Omega t \ll 1$, the particles are not affected by the magnetic field and move ballistically. The mean-square displacement in the $x$- and $y$-directions does not differ from the one in the $z$-direction and so the running diffusion coefficients are both $d_{\parallel}(t) = d_{\perp}(t) = v^2 t/3$ or $d_{\parallel} /(v \Lc) = d_{\perp} /(v \Lc) = (1/3) \Omega t \, \rg/\Lc$.
\item For $1 \ll \Omega t \ll \Omega \taus$, particle motion does get affected by the effective background magnetic field and particles start gyrating about the effective background field direction. This is best seen at low turbulence levels, e.g.\ $\eta = 0.1$ or $0.2$. There is also a suppression of the perpendicular running diffusion coefficient, but in the opposite direction: Whereas for $d_{\parallel}$, the suppression was stronger for larger turbulence level $\eta$, for $d_{\perp}$ the suppression is larger for small $\eta$. This can be understood in the following way: In the limit $\delta B \to 0$, the motion in the $x$- and $y$-directions would be merely a gyration. A finite $\delta B$ misaligns the effective $B$-field direction from the $z$-direction such that part of the gyration in the perpendicular plane now contributes to the parallel motion; in turn, some of the parallel ballistic motion, contributes to the transport in the $x$- and $y$-directions. The larger $\eta$, the stronger is this effect. 
\item For $\Omega \taus \ll \Omega t \ll \Omega \tauc$, perpendicular transport is sub-diffusive, $d_{\perp} \propto t^{\alpha-1}$ with $\alpha < 1$. A similar behaviour had already been observed for non-relativistic particles~\citep{2016MNRAS.459.3395P}. We note that this is reminiscent of so-called compound subdiffusion~\citep{2006ApJ...651..211W}, an effect due to diffusive particle transport along diffusive field lines. However, while compound subdiffusion predicts $d_{\perp} \propto t^{1/2}$, we do not find this scaling for all $\eta$, see the dotted line in Fig.~\ref{fig:dperp_eta}. We note that the transition to this regime sets in later for smaller $\eta$ as $\taus$ is larger.
\item For $\Omega\tauc \ll \Omega t$, finally, perpendicular transport also is diffusive and $d_{\perp}$ attains its asymptotic value, $\kappa_{\perp}$.
\end{enumerate}

\begin{figure*}[tbh]
\includegraphics[scale=1]{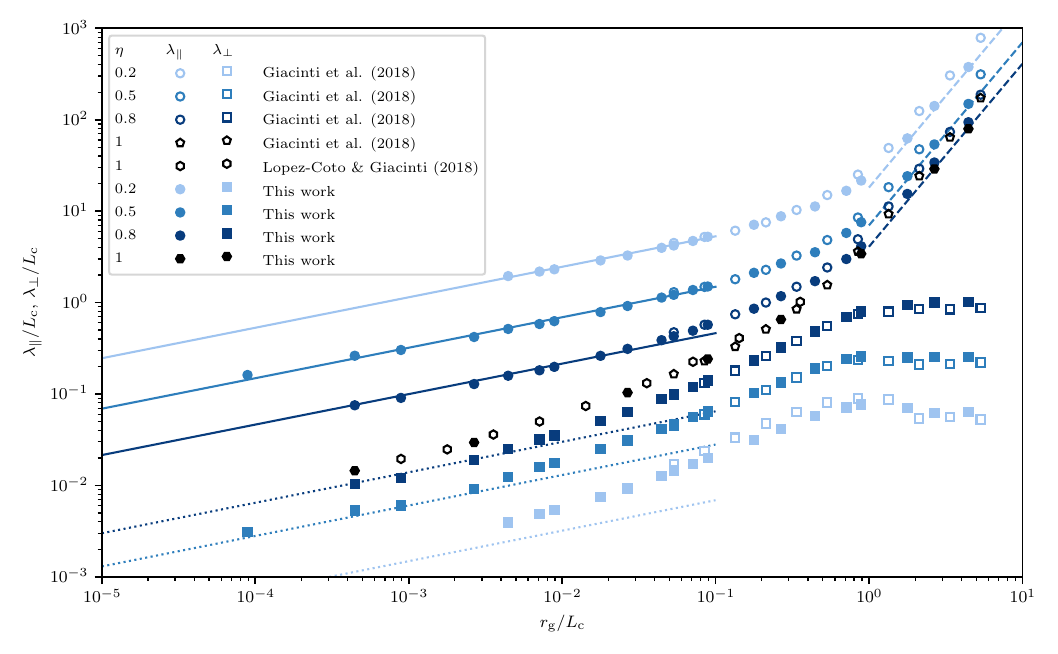}
\caption{
Asymptotic parallel and perpendicular mean free paths $\lambda_{\parallel}$ and $\lambda_{\perp}$ as a function of reduced rigidity $\rg/\Lc$ for different turbulence levels $\eta=0.2, \, 0.5, \, 0.8$ and $1$. We also compare the results of our test particle simulations with the data of \citet{giacinti2017} (empty circles, squares, and pentagons) as well as \citet{Lopez-Coto:2017pbk} (empty hexagons). We have indicated power laws $\propto (\rg/\Lc)^{1/3}$ for $\rg/\Lc < 10^{-1}$ and $\propto (\rg/\Lc)^2$ for $\rg/\Lc > 1$.
}
\label{fig:lams}
\end{figure*}

\begin{figure}[bth]
\includegraphics[width=\columnwidth]{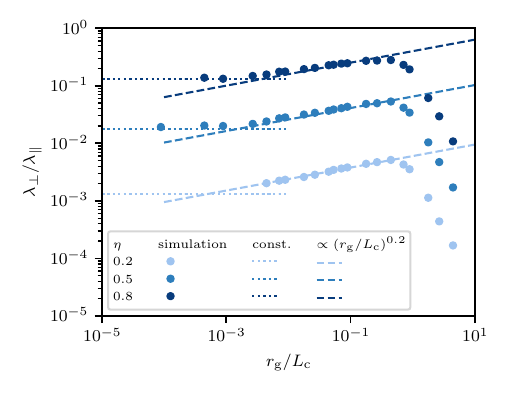}
\caption{
Ratio $\lambda_{\perp} / \lambda_{\parallel}$ of the asymptotic perpendicular and parallel mean free paths as a function of reduced rigidity $\rg/\Lc$ for different turbulence levels $\eta$. We have indicated a constant ratio at low rigidities and a $(\rg/\Lc)^{0.2}$ dependence at intermediate rigidities, as suggested by \citet{2020PhRvD.102j3016D}.
}
\label{fig:ratio}
\end{figure}

In Fig.~\ref{fig:dperp_rigidities} we show the running perpendicular diffusion coefficient $d_{\perp}(t)$, see eq.~\eqref{eqn:def_dperp}, for different reduced rigidities $\rg / \Lc$ and fixed turbulence level $\eta = 0.5$. For $\rg/\Lc \lesssim 1$, the behaviour is qualitatively the same as described above for the case of $\rg / \Lc=8.9 \times 10^{-3}$. For larger values of $\rg/\Lc$, the behaviour becomes markedly different. Here, we can at most make out three phases:
\begin{enumerate}
\item For $\Omega t \ll 1$, the particles move ballistically in the same way as for lower rigidities. 
\item For $1 \ll \Omega t \ll \Omega \taus$, particles gyrate around the background field direction. The direction of the effective background field does not play any role anymore since the particles at these rigidities cannot be in resonance with any of the turbulence modes anymore. Therefore, particles in different realisations gyrate around the original $z$-direction. These gyrations are clearly visible in Fig.~\ref{fig:dperp_rigidities}. 
\item For $\Omega t \gg \Omega \taus$, the gyrations get damped as particles perform small-angle scattering and the asymptotic perpendicular diffusion coefficient $\kappa_{\perp}$ is attained. Note that the value becomes independent of the reduced rigidity $\rg/\Lc$. 
\end{enumerate}

Having established the subdiffusive phase in the perpendicular diffusion coefficient in Fig.~\ref{fig:dperp_eta}, we can revisit Fig.~\ref{fig:dpar_eta} and also identify a subdiffusive phase in the parallel diffusion coefficient, see $d_{\parallel}$ for $\eta = 0.8$ and $100 \lesssim \Omega t \lesssim 1000$. While this might sound surprising, this is unavoidable at high enough turbulence levels; after all in the limit $\eta \to 1$, both running diffusion coefficients must agree. 

Fig.~\ref{fig:dpar_dperp_rigidities} compares the running parallel and perpendicular diffusion coefficients $d_{\parallel}$ and $d_{\perp}$ of individual rigidities in the same panels. While the data presented is the same as in Figs.~\ref{fig:dpar_rigidities} and \ref{fig:dperp_rigidities}, Fig.~\ref{fig:dpar_dperp_rigidities} allows comparing $d_{\parallel}$ and $d_{\perp}$ more directly. In particular this allows directly identifying the time where $d_{\perp}$ starts to be sub-diffusive (for $\rg/\Lc \ll 1$) and where the gyrations in $d_{\perp}$ get damped (for $\rg/\Lc \gtrsim 1$) with the scattering time $\taus'$.

\subsection{Diffusion coefficients as function of rigidity\label{sec:rigidity-dependence}}

\begin{figure}[bht]
\includegraphics[scale=1]{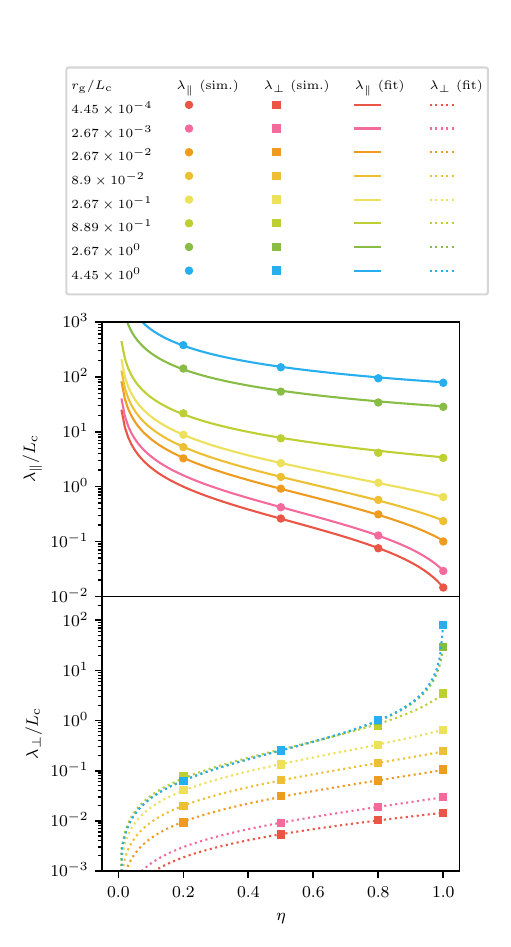}
\caption{
Asymptotic parallel and perpendicular mean free paths $\lambda_{\parallel}$ and $\lambda_{\perp}$ as a function of turbulence levels $\eta$ for a subset of reduced rigidities $\rg/\Lc$. 
The solid and dotted lines show the fitted $\lambda_{\parallel}$ and $\lambda_{\perp}$, respectively, see eqs.~\eqref{eqn:lambda_par_fit} and \ref{eqn:lambda_perp_fit}.}
\label{fig:lams_eta}
\end{figure}

In Fig.~\ref{fig:lams} we show the mean free paths $\lambda_{\parallel}$ and $\lambda_{\perp}$ as functions of rigidity, for the different turbulence levels $\eta = 0.2, \, 0.5, \, 0.8$ and $1$. Note that we have derived the asymptotic values by averaging the values $d_\parallel$ and $d_\perp$ within $\Omega t_{\rm max}/2 \lesssim \Omega t\leq \Omega t_{\rm max}$ where $t_{\rm max}$ is the maximum run time for the simulations. The errors of these estimates are also obtained by averaging the standard errors of the mean within this time period. We have also plotted some lines with constant power law indices. For reduced rigidities $\rg/\Lc \ll 1$, the asymptotic parallel diffusion coefficient $\kappa_{\parallel}$ exhibits the $(\rg/\Lc)^{1/3}$-dependence expected for gyro-resonant interactions due to turbulence with a power spectrum $g(k) \propto k^{-5/3}$. At rigidities $\rg/\Lc \gg 1$, the $(\rg/\Lc)^2$-dependence of small-angle scattering is visible (e.g.~\citep{Gruzinov:2018yxz}). We note that the normalisation of the diffusion coefficient decreases with increasing turbulence level $\eta$. We have also plotted the results from \citet{giacinti2017} which uses the same turbulence model. The agreement is excellent in the range where the simulation data overlap. Note that \citet{2016ApJS..225...20S} has considered rigidities as low as $\sim 2 \times 10^{-2}$; however, the turbulence spectrum considered differed, so we cannot compare directly. By and large, they are in agreement though. 

As far as the perpendicular diffusion coefficient is concerned, its rigidity-dependence is more complicated. Again, we show a power law $\propto (\rg/\Lc)^{1/3}$ that matches the simulation results at low rigidities. However, it appears that the data follow this dependence only over a limited rigidity range. At intermediate rigidities, $\kappa_{\perp}$ grows faster than $(\rg/\Lc)^{1/3}$; for $\rg/\Lc \gtrsim 1$, $\kappa_{\perp}$ becomes constant. The behaviour at intermediate and large rigidities is in line with the behaviour seen in \citet{2020PhRvD.102j3016D}. The normalisation at low rigidities shows the expected ordering in that the diffusion coefficient is smaller for smaller $\eta$. The deviation from the $(\rg/\Lc)^{1/3}$-behaviour sets in at rigidities $\rg/\Lc \sim 10^{-2}$ for $\eta = 0.8$ and at $\rg/\Lc \sim 3 \times 10^{-3}$ for $\eta = 0.5$. We have not actually observed the $(\rg/\Lc)^{1/3}$-scaling for $\eta=0.2$ for the rigidities for which we were able to run simulations. (For low turbulence levels the scattering time $\taus'$ for which $d_{\parallel}$ becomes constant and the time $\tauc$ for which $d_{\perp}$ become constant become prohibitively large.) Overall, the agreement with \citet{giacinti2017} is good. 

\begin{turnpage}
\begin{table*}[tbh]
\begin{tabular}{| l | c | c | c | c | c | c | }
\hline
\multirow{2}{*}{$\rg / \Lc$} & \multicolumn{2}{c |}{$\eta = 0.2$} & \multicolumn{2}{c |}{$\eta = 0.5$} & \multicolumn{2}{c |}{$\eta = 0.8$} \\
\cline{2-7}
& $\lambda_\parallel\left[L_c\right]$ & $\lambda_\perp\left[L_c\right]$ & $\lambda_\parallel\left[L_c\right]$ & $\lambda_\perp\left[L_c\right]$ & $\lambda_\parallel\left[L_c\right]$ & $\lambda_\perp\left[L_c\right]$ \\
\hline
$ 8.90\times 10^{-5} $ & $-$                                         & $-$                                          & $\left(1.614 \pm 0.001\right)\times 10^{-1}$ & $\left(3.096 \pm 0.006\right)\times 10^{-3}$ & $-$                                          & $-$                                          \\
$ 4.45\times 10^{-4} $ & $-$                                         & $-$                                          & $\left(2.621 \pm 0.009\right)\times 10^{-1}$ & $\left(5.331 \pm 0.045\right)\times 10^{-3}$ & $\left(7.527 \pm 0.018\right)\times 10^{-2}$ & $\left(1.040 \pm 0.008\right)\times 10^{-2}$ \\
$ 8.90\times 10^{-4} $ & $-$                                         & $-$                                          & $\left(3.030 \pm 0.001\right)\times 10^{-1}$ & $\left(6.057 \pm 0.048\right)\times 10^{-3}$ & $\left(9.070 \pm 0.011\right)\times 10^{-2}$ & $\left(1.202 \pm 0.001\right)\times 10^{-2}$ \\
$ 2.67\times 10^{-3} $ & $-$                                         & $-$                                          & $\left(4.203 \pm 0.002\right)\times 10^{-1}$ & $\left(9.156 \pm 0.039\right)\times 10^{-3}$ & $\left(1.285 \pm 0.001\right)\times 10^{-1}$ & $\left(1.904 \pm 0.001\right)\times 10^{-2}$ \\
$ 4.45\times 10^{-3} $ & $\left(1.947 \pm 0.050\right)\times 10^{0}$ & $\left(3.945 \pm 0.060\right)\times 10^{-3}$ & $\left(5.148 \pm 0.118\right)\times 10^{-1}$ & $\left(1.234 \pm 0.009\right)\times 10^{-2}$ & $\left(1.584 \pm 0.020\right)\times 10^{-1}$ & $\left(2.486 \pm 0.033\right)\times 10^{-2}$ \\
$ 7.12\times 10^{-3} $ & $\left(2.181 \pm 0.014\right)\times 10^{0}$ & $\left(4.890 \pm 0.048\right)\times 10^{-3}$ & $\left(5.827 \pm 0.052\right)\times 10^{-1}$ & $\left(1.584 \pm 0.010\right)\times 10^{-2}$ & $\left(1.817 \pm 0.022\right)\times 10^{-1}$ & $\left(3.191 \pm 0.013\right)\times 10^{-2}$ \\
$ 8.90\times 10^{-3} $ & $\left(2.310 \pm 0.019\right)\times 10^{0}$ & $\left(5.367 \pm 0.054\right)\times 10^{-3}$ & $\left(6.265 \pm 0.046\right)\times 10^{-1}$ & $\left(1.767 \pm 0.012\right)\times 10^{-2}$ & $\left(1.980 \pm 0.016\right)\times 10^{-1}$ & $\left(3.499 \pm 0.024\right)\times 10^{-2}$ \\
$ 1.78\times 10^{-2} $ & $\left(2.890 \pm 0.026\right)\times 10^{0}$ & $\left(7.503 \pm 0.057\right)\times 10^{-3}$ & $\left(7.872 \pm 0.061\right)\times 10^{-1}$ & $\left(2.489 \pm 0.016\right)\times 10^{-2}$ & $\left(2.613 \pm 0.020\right)\times 10^{-1}$ & $\left(5.092 \pm 0.018\right)\times 10^{-2}$ \\
$ 2.67\times 10^{-2} $ & $\left(3.270 \pm 0.025\right)\times 10^{0}$ & $\left(9.303 \pm 0.048\right)\times 10^{-3}$ & $\left(9.172 \pm 0.096\right)\times 10^{-1}$ & $\left(3.111 \pm 0.017\right)\times 10^{-2}$ & $\left(3.116 \pm 0.022\right)\times 10^{-1}$ & $\left(6.378 \pm 0.043\right)\times 10^{-2}$ \\
$ 4.45\times 10^{-2} $ & $\left(3.968 \pm 0.036\right)\times 10^{0}$ & $\left(1.275 \pm 0.005\right)\times 10^{-2}$ & $\left(1.133 \pm 0.008\right)\times 10^{0}$  & $\left(4.158 \pm 0.024\right)\times 10^{-2}$ & $\left(3.879 \pm 0.020\right)\times 10^{-1}$ & $\left(8.822 \pm 0.065\right)\times 10^{-2}$ \\
$ 5.34\times 10^{-2} $ & $\left(4.208 \pm 0.028\right)\times 10^{0}$ & $\left(1.453 \pm 0.007\right)\times 10^{-2}$ & $\left(1.215 \pm 0.007\right)\times 10^{0}$  & $\left(4.703 \pm 0.021\right)\times 10^{-2}$ & $\left(4.282 \pm 0.024\right)\times 10^{-1}$ & $\left(9.951 \pm 0.055\right)\times 10^{-2}$ \\
$ 7.12\times 10^{-2} $ & $\left(4.718 \pm 0.025\right)\times 10^{0}$ & $\left(1.721 \pm 0.008\right)\times 10^{-2}$ & $\left(1.373 \pm 0.015\right)\times 10^{0}$  & $\left(5.596 \pm 0.047\right)\times 10^{-2}$ & $\left(4.925 \pm 0.030\right)\times 10^{-1}$ & $\left(1.196 \pm 0.006\right)\times 10^{-1}$ \\
$ 8.90\times 10^{-2} $ & $\left(5.245 \pm 0.030\right)\times 10^{0}$ & $\left(1.997 \pm 0.013\right)\times 10^{-2}$ & $\left(1.499 \pm 0.011\right)\times 10^{0}$  & $\left(6.467 \pm 0.041\right)\times 10^{-2}$ & $\left(5.723 \pm 0.028\right)\times 10^{-1}$ & $\left(1.410 \pm 0.006\right)\times 10^{-1}$ \\
$ 1.78\times 10^{-1} $ & $\left(7.097 \pm 0.061\right)\times 10^{0}$ & $\left(3.125 \pm 0.021\right)\times 10^{-2}$ & $\left(2.116 \pm 0.011\right)\times 10^{0}$  & $\left(1.024 \pm 0.006\right)\times 10^{-1}$ & $\left(8.572 \pm 0.058\right)\times 10^{-1}$ & $\left(2.329 \pm 0.012\right)\times 10^{-1}$ \\
$ 2.67\times 10^{-1} $ & $\left(8.781 \pm 0.067\right)\times 10^{0}$ & $\left(4.131 \pm 0.015\right)\times 10^{-2}$ & $\left(2.676 \pm 0.025\right)\times 10^{0}$  & $\left(1.331 \pm 0.004\right)\times 10^{-1}$ & $\left(1.175 \pm 0.005\right)\times 10^{0}$  & $\left(3.225 \pm 0.016\right)\times 10^{-1}$ \\
$ 4.45\times 10^{-1} $ & $\left(1.126 \pm 0.009\right)\times 10^{1}$ & $\left(5.784 \pm 0.025\right)\times 10^{-2}$ & $\left(3.551 \pm 0.038\right)\times 10^{0}$  & $\left(1.893 \pm 0.010\right)\times 10^{-1}$ & $\left(1.715 \pm 0.010\right)\times 10^{0}$  & $\left(4.813 \pm 0.020\right)\times 10^{-1}$ \\
$ 7.12\times 10^{-1} $ & $\left(1.668 \pm 0.013\right)\times 10^{1}$ & $\left(7.138 \pm 0.037\right)\times 10^{-2}$ & $\left(5.778 \pm 0.043\right)\times 10^{0}$  & $\left(2.405 \pm 0.014\right)\times 10^{-1}$ & $\left(2.992 \pm 0.015\right)\times 10^{0}$  & $\left(6.917 \pm 0.022\right)\times 10^{-1}$ \\
$ 8.90\times 10^{-1} $ & $\left(2.159 \pm 0.016\right)\times 10^{1}$ & $\left(7.651 \pm 0.048\right)\times 10^{-2}$ & $\left(7.562 \pm 0.055\right)\times 10^{0}$  & $\left(2.582 \pm 0.016\right)\times 10^{-1}$ & $\left(4.129 \pm 0.028\right)\times 10^{0}$  & $\left(7.969 \pm 0.038\right)\times 10^{-1}$ \\
$ 1.78\times 10^{0} $ & $\left(6.249 \pm 0.036\right)\times 10^{1}$ & $\left(7.071 \pm 0.037\right)\times 10^{-2}$ & $\left(2.410 \pm 0.017\right)\times 10^{1}$   & $\left(2.501 \pm 0.016\right)\times 10^{-1}$ & $\left(1.543 \pm 0.012\right)\times 10^{1}$  & $\left(9.461 \pm 0.055\right)\times 10^{-1}$ \\
$ 2.67\times 10^{0} $ & $\left(1.413 \pm 0.006\right)\times 10^{2}$ & $\left(6.203 \pm 0.032\right)\times 10^{-2}$ & $\left(5.371 \pm 0.043\right)\times 10^{1}$   & $\left(2.535 \pm 0.014\right)\times 10^{-1}$ & $\left(3.402 \pm 0.017\right)\times 10^{1}$  & $\left(1.002 \pm 0.004\right)\times 10^{0}$  \\
$ 4.45\times 10^{0} $ & $\left(3.768 \pm 0.020\right)\times 10^{2}$ & $\left(6.321 \pm 0.039\right)\times 10^{-2}$ & $\left(1.489 \pm 0.009\right)\times 10^{2}$   & $\left(2.545 \pm 0.014\right)\times 10^{-1}$ & $\left(9.386 \pm 0.056\right)\times 10^{1}$  & $\left(1.013 \pm 0.004\right)\times 10^{0}$  \\
\hline
\end{tabular}
\caption{Asymptotic parallel and perpendicular mean free paths $\lambda_{\parallel}$ and $\lambda_{\perp}$ for different reduced rigidities $\rg/\Lc$ and at the three turbulence levels $\eta = 0.2,\, 0.5,\,{\rm and}\,0.8$.}
\label{tbl1}
\end{table*}
\end{turnpage}

\begin{table*}[thb]
\begin{tabular}{| l | c | c | c |}
\hline
\multirow{2}{*}{$\rg / \Lc$} & $\eta = 1$ & \multirow{2}{*}{$\lambda_{0,\parallel} [\Lc]$} & \multirow{2}{*}{$\lambda_{0,\perp} [\Lc]$}\\
\cline{2-2}
& $\lambda_1 [\Lc]$ & & \\
\hline
$ 4.45\times 10^{-4} $ & $\left(1.449 \pm 0.022\right)\times 10^{-2}$ & $\left(2.457 \pm 0.012\right)\times 10^{-1}$ & $\left(8.571 \pm 0.276\right)\times 10^{-3}$ \\
$ 2.67\times 10^{-3} $ & $\left(2.950 \pm 0.031\right)\times 10^{-2}$ & $\left(3.929 \pm 0.009\right)\times 10^{-1}$ & $\left(1.330 \pm 0.019\right)\times 10^{-2}$ \\
$ 2.67\times 10^{-2} $ & $\left(1.035 \pm 0.007\right)\times 10^{-1}$ & $\left(8.077 \pm 0.089\right)\times 10^{-1}$ & $\left(4.216 \pm 0.040\right)\times 10^{-2}$ \\
$ 8.90\times 10^{-2} $ & $\left(2.418 \pm 0.016\right)\times 10^{-1}$ & $\left(1.265 \pm 0.011\right)\times 10^{0}$ & $\left(8.721 \pm 0.089\right)\times 10^{-2}$ \\
$ 2.67\times 10^{-1} $ & $\left(6.535 \pm 0.038\right)\times 10^{-1}$ & $\left(2.035 \pm 0.024\right)\times 10^{0}$ & $\left(1.706 \pm 0.011\right)\times 10^{-1}$ \\
$ 8.90\times 10^{-1} $ & $\left(3.415 \pm 0.021\right)\times 10^{0}$ & $\left(4.329 \pm 0.060\right)\times 10^{0}$ & $\left(2.875 \pm 0.021\right)\times 10^{-1}$ \\
$ 2.67\times 10^{0} $ & $\left(2.896 \pm 0.011\right)\times 10^{1}$ & $\left(2.687 \pm 0.031\right)\times 10^{1}$ & $\left(2.545 \pm 0.012\right)\times 10^{-1}$ \\
$ 4.45\times 10^{0} $ & $\left(7.944 \pm 0.020\right)\times 10^{1}$ & $\left(7.240 \pm 0.077\right)\times 10^{1}$ & $\left(2.550 \pm 0.013\right)\times 10^{-1}$ \\
\hline
\end{tabular}
\caption{Asymptotic mean free paths $\lambda_1$ for different reduced rigidities $\rg/\Lc$ at $\eta = 1$ and the fit parameters $\lambda_{0,\parallel}$ and $\lambda_{0,\perp}$ for eqs. \ref{eqn:lambda_par_fit} and \ref{eqn:lambda_perp_fit}.}
\label{tbl2}
\end{table*}

In Fig.~\ref{fig:ratio} we show the ratio of the perpendicular and parallel mean free paths, $\lambda_{\perp} / \lambda_{\parallel}$ as functions of rigidity, for the different turbulence levels $\eta = 0.2, 0.5$ and $0.8$. In the range between $\rg/\Lc \simeq 10^{-2}$ and $1$, the ratio is increasing with a behaviour close to the $(\rg/\Lc)^{0.2}$ suggested by \citet{2020PhRvD.102j3016D}. This is due to the faster than $(\rg/\Lc)^{1/3}$-dependence of $\lambda_{\perp}$ at these rigidities. At lower rigidities, however, the ratio becomes constant, meaning that both $\lambda_{\parallel}$ and $\lambda_{\perp}$ scale in the same way. At rigidities $\gtrsim 1$, the ratio drops steeply due to fact that $\lambda_{\perp}$ becomes constant while $\lambda_{\parallel} \propto (\rg/\Lc)^2$. 

In Fig.~\ref{fig:lams_eta}, we have plotted the asymptotic parallel and perpendicular mean free paths $\lambda_{\parallel}$ and $\lambda_{\perp}$ as a function of turbulence level $\eta$ for a subset of the reduced rigidities. We have also investigated the scaling with $\delta B^2 / B_0^2$ that had been anticipated at small turbulence levels, see eqs.~\eqref{eqn:lambda_par_generic} and ~\eqref{eqn:lambda_perp_generic}. As $\eta \to 1$, of course, both $\lambda_{\parallel}$ and $\lambda_{\perp}$ must coincide, $\lambda_{\parallel}, \lambda_{\perp} \to \lambda_1$. A simple way to achieve this scaling is to extend the ($\delta B^2 / B_0^2$)-dependence of eqs.~\eqref{eqn:lambda_par_generic} and ~\eqref{eqn:lambda_perp_generic} to
\begin{align}
\lambda_{\parallel} &= \left( \lambda_1 + \lambda_{0,\parallel} \left( \frac{\delta B^2}{B_0^2} \right)^{-1} \right) \, , \label{eqn:lambda_par_fit} \\
\lambda_{\perp} &= \left( \lambda_1^{-1} + \lambda_{0,\perp}^{-1} \left( \frac{\delta B^2}{B_0^2} \right)^{-1} \right)^{-1} \, . \label{eqn:lambda_perp_fit}
\end{align}
For those reduced rigidities, for which we have simulation data at the turbulence level $\eta = 1$, we know $\lambda_1$ and hence, we have only the free parameters $\lambda_{0,\parallel}$ and $\lambda_{0,\perp}$ for each reduced rigidity. We have fitted eqs.~\eqref{eqn:lambda_par_fit} and \eqref{eqn:lambda_perp_fit} to the simulation data and show the resulting fits in Fig.~\ref{fig:lams_eta}. We find that eqs.~\eqref{eqn:lambda_par_fit} and \eqref{eqn:lambda_perp_fit} give an excellent representation of the data. 

Our asymptotic mean free paths are provided in Tbl.~\ref{tbl1} (for $\eta = 0.2, 0.5$ and $0.8$) and in Tbl.~\ref{tbl2} (for $\eta = 1$). We note that eqs.~\eqref{eqn:lambda_par_fit} and \eqref{eqn:lambda_perp_fit} attempt to provide a general expression for $\lambda_\parallel$ and $\lambda_\perp$, which are functions of both $\delta B^2/B_0^2$ and $r_g/L_c$. The dependence on reduced rigidity of the mean free paths is, roughly speaking, factorized into $\lambda_1$, $\lambda_{0,\parallel}$, and $\lambda_{0,\perp}$ which depends only on $r_g/L_c$ (as shown in Tbl.~\ref{tbl2}).

\subsection{Field line running diffusion coefficients}

\begin{figure}[!thb]
\includegraphics[scale=1]{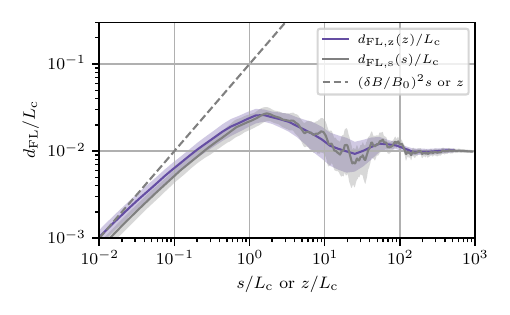}
\caption{Running field line diffusion coefficients $d_{\text{FL}}(s)$ with respect to arc length $s$ (purple) and $d_{\text{FL}}(z)$ with respect to the $z$-coordinate (gray) for a turbulence level $\eta = 0.091$. The lines show the mean over the ensemble of turbulent magnetic fields and the shaded bands indicate the standard mean error.}
\label{fig:deriv_d_FL_eta_0.091}
\end{figure}

\begin{figure}[!thb]
\includegraphics[scale=1]{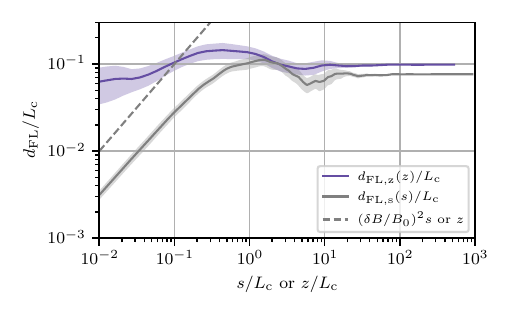}
\caption{Same as Fig.~\ref{fig:deriv_d_FL_eta_0.5}, but for a turbulence level $\eta=0.5$.}
\label{fig:deriv_d_FL_eta_0.5}
\end{figure}

In Fig.~\ref{fig:deriv_d_FL_eta_0.091}, we show the running field line diffusion coefficients $d_{\text{FL},s}(s)$ with respect to arc length $s$ and $d_{\text{FL},z}(z)$ with respect to the $z$-coordinate for a turbulence level $\eta = 0.091$. At this low turbulence level, the field lines are approximately aligned with the $z$-direction and so the difference between the definition with respect to arc length $s$ or $z$-coordinate are very small. The two curves seem to differ mostly in the level of fluctuations, but this is caused by the fact that we needed to smoothen the $d_{\text{FL}}$ defined with respect to $z$ for technical reasons, see Sec.~\ref{sec:FL_diffusion_coeff}. As a function of $s$ and $z$, respectively, both running diffusion coefficients grow close to $\propto s$ and $\propto z$, initially, that is for $s \ll \Lc$ and $z \ll \Lc$. Once the correlation length has been exceeded, the running diffusion coefficients start to grow more slowly. However, the diffusive regime is not attained immediately, but rather via some sub-diffusive phase between $\sim \Lc$ and $\sim 10 \Lc$. The diffusive field-line wandering regime is finally attained for very large $s$ or $z$, that is beyond $\sim 100 \Lc$. We note that a similar subdiffusive phase can be observed in the field-line simulations of \citet{2016ApJS..225...20S} for the case of a small turbulence level, see their Fig.~14. We caution however, that a somewhat different definition of the running field line diffusion coefficient was adopted there, so their values cannot be compared directly with our values. 

In fact, we have provided also a comparison between the numerical results of the running field line diffusion and the ones derived from a semi-analytic model in the supplemental material of a companion paper~\citep{short_paper}. In this case, the running field line diffusion coefficient is evaluated from a set of differential equations as introduced by \citet{2016ApJS..225...20S} and, with this semi-analytic formulation, we find also subdiffusive behaviour in the running field line diffusion coefficient for scales around the correlation length in isotropic turbulence. In fact, the transport of field lines on these small scales depends rather sensitively on the particular type of turbulence considered. For example, the running field line diffusion coefficient is expected to exhibit a direct transition from ballistic on small scales to diffusive on large scales only in slab turbulence \citep{shalchi2021}. It has been argued \citep{lazarian2014}, however, that the field line transport can be superdiffusive on small scales for different types of Alfv\'enic turbulence. Such a superdiffusive behaviour is also predicted for 2D and slab (composite) turbulence \citep{2009ASSL..362.....S}. It has also been pointed out that the transport of field lines on small scales below the correlation length can have implications for theories of CR acceleration~\citep{lazarian2014} and escape around their sources in case where turbulence is not self-generated by CRs~\citep{nava2013,recchia2021}. In other words, the transport of field lines might play an important role in determining radiative signatures induced by CRs around sources and, thus, our understanding on the running field line diffusion coefficient could be employed together with these observations to diagnose turbulence in these systems.

At larger turbulence levels, the differences between the definitions of the field line diffusion coefficients with respect to arc length $s$ and $z$-coordinate become more pronounced. This is visible in Fig.~\ref{fig:deriv_d_FL_eta_0.5}, where we show the running field line diffusion coefficients $d_{\text{FL},s}(s)$ with respect to arc length $s$ and $d_{\text{FL},z}(z)$ with respect to the $z$-coordinate for a turbulence level $\eta = 0.5$. We remind the reader that $\eta = 0.5$ means that the turbulent field $\vct{\delta B}$ has the same energy density as the background field $\vct{B}_0$. Qualitatively, both diffusion coefficients show a similar behaviour as at low turbulence level, in particular the field line transport is first ballistic, then sub-diffusive and eventually diffusive. However, $d_{\text{FL,z}}$ is almost always larger in value than $d_{\text{FL,s}}$. In particular, in the ballistic phase, it is larger by almost an order of magnitude and in the late diffusive phase, $d_{\text{FL,z}} \to \kappa_{\text{FL,z}} \simeq 0.1 \Lc$ while $d_{\text{FL,s}} \to \kappa_{\text{FL,s}} \simeq 0.076 \Lc$. At least qualitatively, this difference can be understood as follows: For finite turbulence level $\eta$, the field lines are not generally aligned with the $z$-direction. Part of the arc length $s$ is therefore accumulated also in the $x$- and $y$-directions and so $\langle (\Delta z)^2 \rangle < (\Delta s)^2 = \langle (\Delta x)^2 + (\Delta y)^2 + (\Delta z)^2 \rangle$. Therefore, $\dd \langle (\Delta x)^2 \rangle / \dd z > \dd \langle (\Delta x)^2 \rangle / \dd s$.

\section{Discussion\label{sec:discussion}}

In the following, we will consider some examples where the difference in scaling of the perpendicular mean free path has important phenomenological consequences. 

Perpendicular transport is also central in shock acceleration at quasi-perpendicular shocks. For supernova remnants (SNRs), there is consensus that magnetic fields need to be amplified, in order for shock-accelerated particles to reach the knee energy. In that situation the turbulence spectrum would likely differ from the Kolmogorov form assumed here. However, this does not need to be the case for all SNRs, see e.g.~\citet{2014ApJ...792..133F}. Interplanetary shocks, e.g. in CMEs, would be another acceleration site where the acceleration rate and hence the maximum energy is regulated by perpendicular transport~\citep{2010AdSpR..46.1208D}. 

In the remainder of this section, we will confine ourselves to considering the implications of our findings for the transport of Galactic CRs in more detail, specifically of the difference in scaling of the perpendicular diffusion coefficient $\kappa_{\perp}$ with reduced rigidity $\rg/\Lc$. Given the relatively small difference of $\sim0.2$ in spectral index, one might wonder what such implications could be. For example, for $\eta=0.5$, if they agree at $\rg/\Lc=3 \times 10^{-3}$ (which corresponds to a rigidity of $300 \, \text{TV}$ for our benchmark values), they differ by about an order of magnitude at $\rg/\Lc=10^{-8}$ (which corresponds to $1 \, \text{GV}$). In the following, we will consider the competition of perpendicular escape from a region of size $L$ with three processes: parallel escape, advection and cooling losses. 

\paragraph{Diffusive escape}

First, we consider competition between parallel and perpendicular escape. To quantify this, we define the rates $t_{\parallel} \equiv L_{\parallel}^2 / (2 \kappa_{\parallel})$ and $t_{\perp} \equiv L_{\perp}^2 / (2 \kappa_{\perp})$. Whether parallel or perpendicular loss dominates depends on the aspect ratio of the system, $L_{\perp}/L_{\parallel}$. Perpendicular escape is dominant if $\kappa_{\perp}/\kappa_{\parallel} > (L_{\perp}/L_{\parallel})^2$. 

For instance for the Galaxy, we estimate $L_{\perp}/L_{\parallel} = 0.1 \mathellipsis 0.3$. Therefore, if the relative scaling had been $\kappa_{\perp}/\kappa_{\parallel} \propto (\rg/\Lc)^{0.2}$ at all rigidities, perpendicular transport could have been important only at the highest rigidities, if at all. However, we have found that $\kappa_{\perp}/\kappa_{\parallel} = \text{const.}$ Specifically, for $\eta=0.5$, $\kappa_{\perp}/\kappa_{\parallel} = 2 \times 10^{-2}$ below $\rg/\Lc = 3 \times 10^{-3}$. 
Therefore, perpendicular escape can be important at all energies below $300 \, \text{TV}$.

\paragraph{Advective escape}

Next, we compare perpendicular diffusive escape with the time scale $t_{\perp} \equiv L_{\perp}^2 / (2 \kappa_{\perp})$ with escape due to advection at the speed $U$. We characterise this process with the time scale $t_{\text{adv}} \equiv L_{\perp} / U$. As $t_{\perp}$ increases with decreasing rigidity $\rg/\Lc$, advection will dominate over diffusion below some critical rigidity $r_{\text{g,a}}/\Lc$ which is defined through $t_{\text{adv}} = t_{\perp}(r_{\text{g,a}}/\Lc)$. Note that $r_{\text{g,a}}/\Lc$ will sensitively depend on whether $\kappa_{\perp} \propto (\rg/\Lc)^{1/3}$ or $(\rg/\Lc)^{1/2}$. 

For a numerical estimate, we fix the perpendicular diffusion coefficient at $\rg/\Lc = 3 \times 10^{-3}$ to $\kappa_{\perp} = (v/3) \lambda_{\perp} = (v/3) 10^{-2} \, \Lc$. We further adopt an advection speed $U = 2 \, \text{km} \, \text{s}^{-1} \simeq (2/3) 10^5 \, v$ with the particle speed $v$ set to $c$ and set $L_{\perp} = 100 \, \Lc$. Solving for $r_{\text{g,a}}/\Lc$, we find $r_{\text{g,a}}/\Lc = 3 \times 10^{-6}$. Had we assumed $\kappa_{\perp} \propto (\rg/\Lc)^{1/2}$, this would be $r_{\text{g,a}}/\Lc = 3 \times 10^{-5}$ instead, that is an order of magnitude larger. 

\paragraph{Cooling losses}

Finally, we consider the transition between transport dominated by perpendicular losses with the loss time $t_{\perp}$ and cooling by synchrotron or Inverse Compton scattering. This is an important phenomenological question when considering the transport of CR electrons and positrons~\citep{Mertsch:2018bqd,Lipari:2018usj}. In the Thomson regime and assuming relativistic particles, the energy loss rate is scaling as $b(E)\sim E^2$ and, thus the cooling time scales like $t_{\text{cool}} \sim E/b(E) \sim (\rg/\Lc)^{-1}$, hence diffusive losses will dominate a low rigidities, cooling losses will dominate at high rigidities. We define the transition to take place at $r_{\text{g,c}}/\Lc$ defined through $t_{\text{cool}}(r_{\text{g,c}}/\Lc) = t_{\perp}(r_{\text{g,c}}/\Lc)$. We adopt some benchmark numbers for a numerical estimate, that is we demand $t_{\text{cool}}(\mathcal{R} = 1 \, \text{GV}) = 3 \times 10^{16} \, \text{s}$ that is $t_{\text{cool}}(\rg/\Lc) = 10^{7} \Lc / c \left[(\rg/\Lc)/10^{-8}\right]^{-1}$ for our benchmark magnetic field parameters. Thus for the $\kappa_{\perp} \propto (\rg/\Lc)^{1/2}$ scaling, the transition takes place at $r_{\text{g,c}}/\Lc = 1.5 \times 10^{-12}$ whereas $r_{\text{g,c}}/\Lc = 3.1 \times 10^{-10}$ for a $\kappa_{\perp} \propto (\rg/\Lc)^{1/3}$ scaling, a difference of a factor $\sim 200$. 

\section{Conclusion\label{sec:conclusion}}

Perpendicular transport of high-energy particles is important in a number of environments and a sound theoretical understanding is important when interpreting observations, be it of \textit{in-situ} observations in the heliosphere, studies of Galactic CRs or non-thermal emission from sources. Our simulations have shown that the rigidity-dependence of the perpendicular mean free path $\lambda_{\perp}$ differs from that of the parallel mean free path $\lambda_{\parallel}$ for reduced rigidities $\rg/\Lc \lesssim 1$; at even lower reduced rigidities $\rg/\Lc \ll 1$, however, the perpendicular diffusion coefficient returns back to the same scaling. Specifically, for Kolmogorov turbulence, $\lambda_{\perp} \propto (\rg/\Lc)^{0.5}$ for $\rg/\Lc \lesssim 1$ and $\propto (\rg/\Lc)^{1/3}$ for $\rg/\Lc \ll 1$ whereas $\lambda_{\parallel} \propto (\rg/\Lc)^{1/3}$ for all $\rg/\Lc \lesssim 1$. Previous analyses had instead speculated about the $\lambda_{\perp} \propto (\rg/\Lc)^{0.5}$ behaviour extending to the lowest rigidities. In a companion paper~\citep{short_paper}, we have provided an analytical model that is able to reproduce this scaling as long as the subdiffusive phase in the running field line diffusion coefficient is taken into account. 

\vspace{2em}

The authors would like to thank Carmelo Evoli for helpful discussions. This project was funded by the Deutsche Forschungsgemeinschaft (DFG, German Research Foundation) -- project number 426614101.

\bibliography{DiffIsoTurb}

\appendix

\section{Systematic uncertainties\label{sec:uncertainties}}

In the following, we present some results obtained by varying the methodology used in the test particle simulations and in the computation of the diffusion coefficients. 

\subsection{Harmonic vs. Grid Method}
\label{sec:harmonic_vs_grid}

\begin{figure*}
\includegraphics[scale=1]{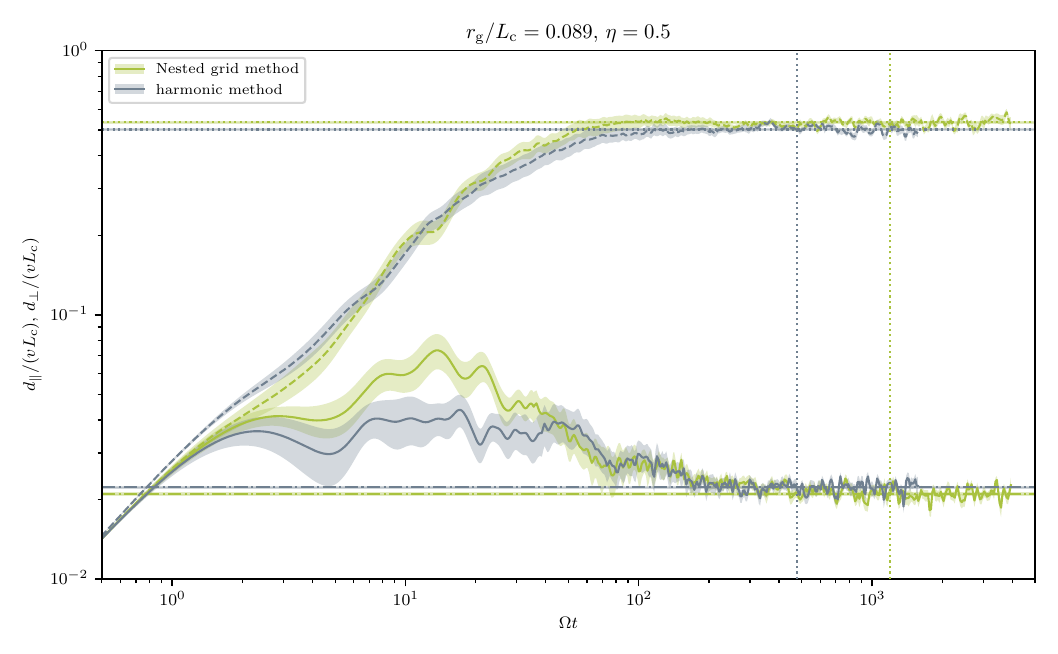}
\caption{
Comparison of the running diffusion coefficients using the harmonic and the grid-based methods for setting up the turbulent magnetic fields for $\rg/\Lc = 0.089$ and $\eta = 0.5$. The parallel (perpendicular) running diffusion coefficient are indicated by dashed (solid) lines, their asymptotic values by dotted (dot-dashed) lines; shaded bands indicate the respective standard mean errors. The asymptotic values have been computed by averages beyond the time indicated by the dotted vertical lines. 
}%
\label{fig:harmonic-grid-10PeV}
\includegraphics[scale=1]{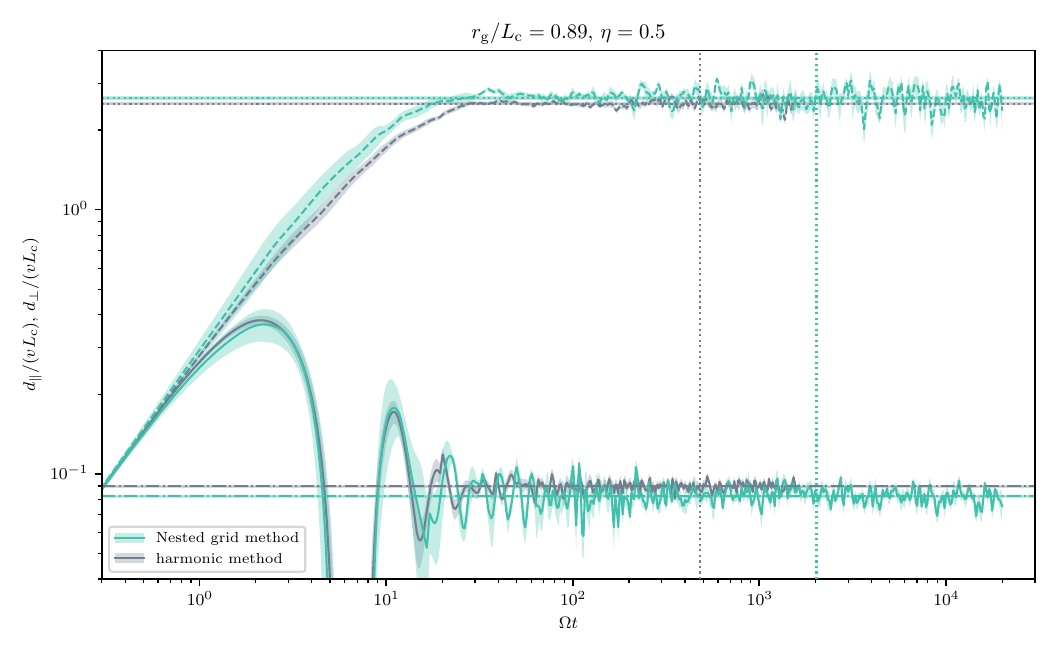}
\caption{Same as Fig.~\ref{fig:harmonic-grid-10PeV}, but for $\rg/\Lc = 0.89$.}
\label{fig:harmonic-grid}
\end{figure*}

In Sec.~\ref{sec:synthetic_turbulence}, we reviewed two methods for setting up the turbulent magnetic field in test particle simulations: the harmonic method and the nested grid method, the latter of which we employed in our work. As discussed, only the nested grid method allows covering the dynamical range needed to simulate particles at $\text{TV}$ rigidities, but it requires periodically replicating the various grids in order to cover the distances that particles travel during the simulations. We have attempted to prevent the introduction of possible artefacts due to this periodicity effect by allowing for generous amounts of padding. Yet, it is desirable to compare the results from the nested grid method with those from the harmonic method. Possible deviations could be interpreted as a sign of systematic issues in either of the methods. 

However, given the computational expense of the harmonic method, such a comparison is only possible at relatively large rigidities. Here, we perform said check at reduced rigidities of $\rg/\Lc = 0.089$ and $0.89$. While this is far from the low rigidities that are accessible in the nested grid method, $\rg/\Lc = 0.089$ is in a range of particular interest where the rigidity-dependence of the perpendicular diffusion coefficient deviates from that of the parallel one, see, e.g.\ Fig.~\ref{fig:ratio}. Furthermore, at $\rg/\Lc = 0.089$, particles interact resonantly with turbulent waves as witnessed by the rigidity-scaling of $\lambda_{\parallel}$ close to $\propto (\rg/\Lc)^{1/3}$, see Fig.~\ref{fig:lams}. 

For the nested grid method, the same grid configurations as discussed in Sec.~\ref{sec:nested_grid_method} were used. For the harmonic method, we have logarithmically spaced $10^5$ modes between $k_{\text{min}} = 2 \pi / L_{\text{max}}$ and $k_{\text{max}} = 10^5 k_{\text{min}}$. In Figs.~\ref{fig:harmonic-grid-10PeV} and \ref{fig:harmonic-grid} we compare the running parallel and perpendicular diffusion coefficients. 

The running diffusion coefficients show the same time-dependences and overall they agree very well: At early times, $\Omega t \ll 1$, the running diffusion coefficients grow ballistically. The parallel running diffusion coefficient transitions to a diffusive behaviour beyond the diffusion time $\taus'$. The perpendicular running diffusion coefficient instead experiences some subdiffusive behaviour before also becoming diffusive for $t \gg \tauc$. Note that for $\rg/\Lc = 0.89$ there are some features related to gyrations of particles around the effective background field visible. The differences between the running diffusion coefficients obtained with both methods almost always remain within the uncertainties introduced by the sample variance which in Figs.~\ref{fig:harmonic-grid-10PeV} and \ref{fig:harmonic-grid} are indicated by the shaded bands. The differences in the asymptotic diffusion coefficients are $6 \, \%$ and $7 \, \%$ for $\rg/\Lc = 0.089$ and $\rg/\Lc = 0.89$, respectively, which is larger than the uncertainties since the asymptotic diffusion coefficients are based on an average of the running diffusion coefficients over a large number of time steps. Unfortunately, we have not been able to reach better convergence of both methods and so we consider this as an estimate of the systematic uncertainty of our test particle simulations. 
However, for the purpose of this analysis, this level of accuracy is certainly sufficient.

Note that we expect periodicity effects, if present, to be even weaker at smaller rigidities since for the maximum runtimes considered in our simulations, particles generally travel over smaller distances. We thus conclude that it is very unlikely that our results are affected by periodicity effects due to the nested grid method.

\subsection{Differential Equation Solvers\label{sec:test_solvers}}

\begin{figure}
\centering
\includegraphics[scale=1]{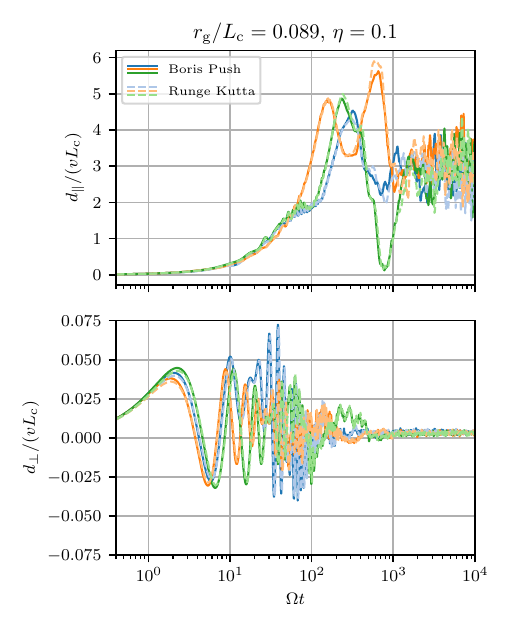}
\caption{
Comparison of the running diffusion coefficients computed with the Boris-Push solver and an adaptive fourth-order Runge-Kutta solver. We have adopted a reduced rigidity of $\rg/\Lc = 0.089$ and the turbulence level $\eta = 0.1$. In the top panel, the running parallel diffusion coefficient $d_{\parallel}$ is shown, in the bottom panel the running parallel diffusion coefficient $d_{\perp}$. The colours blue, orange and green refer to three turbulent field realisations, with the results for the Boris-Push method shown with solid lines and dark colours and those for the Runge-Kutta solve with dashed lines and light colours. Note that the same random realisations have been used for the Boris-Push and the Runge-Kutta methods. We have used a linear ordinate axis in order to be able to appreciate the similarity between the results for both methods. 
}
\label{fig:Borisrk}
\end{figure}

Another potential source of systematic errors in test particle simulations is the numerical solver for the equations of motion. In the literature, a number of solvers is being employed, as discussed in Sec.~\ref{sec:solvers} and we have used the Boris-Push method for all our results so far. Here, we compare the results for running diffusion coefficients for one combination of rigidity and turbulence level obtained with the Boris-Push method and with a fourth-order Runge-Kutta solver. 

Generally speaking, the adaptive Runge-Kutta method is computationally significantly more expensive, in particular at small rigidities. We have therefore performed the comparison at a reduced rigidity of $\rg/\Lc = 0.089$ much larger than what can be achieved with the Boris-Push method alone. While we have used the same realisations of the turbulent magnetic field for both methods, the accumulation of numerical errors leads to the divergence of individual particle trajectories with time even if the same initial conditions are chosen. In order to delay this divergence of trajectories, we have also chosen a rather low turbulence level of $\eta = 0.1$. 

In Fig.~\ref{fig:Borisrk}, we show the results for the running parallel and perpendicular diffusion coefficients, $d_{\parallel}$ and $d_{\perp}$ as a function of time. The colours blue, orange and green refer to three different random realisations of the turbulent magnetic field. The results for the Boris-Push method are shown with solid lines and dark colours, those for the Runge-Kutta method with dashed lines and light colours. We have chosen a linear ordinate axis in order to better compare the results of both methods for individual realisations of the turbulent field. 

The running diffusion coefficients in both methods largely show the same behaviour. While individual trajectories differ, as discussed above, the running diffusion coefficients agree rather well. Also, the running diffusion coefficients in \emph{different} realisations of the turbulent magnetic field agree at early times (in the ballistic phase) and at late times (modulo the numerical noise). At intermediate times, they exhibit the largest sensitivity to the particular realisation of the turbulent magnetic field. 

Besides the computational expense, the Runge-Kutta method also does not conserve particle energy. We know, however, that in the magnetostatic setup, particle energy is conserved. We have checked the loss of particle energy in the Runge-Kutta method by tracking the average velocity magnitude in units of the speed of light, $|\vct{v}|/c$. For low reduced rigidity of $\rg/\Lc = 8.9 \times 10^{-4}$ and for $\eta = 0.5$ we need to run for several hundred thousands of gyro times in order for particle transport to become diffusive. For these input values, we have found that $|\vct{v}|/c$ has on average decreased by more than $15 \, \%$ for $\Omega t = 10^5$ already. Note that in the Boris-Push method, particle energy is conserved by construction. 

In conclusion, we have confirmed that the Boris-Push and the Runge-Kutta methods give approximately the same running diffusion coefficients at intermediate particle rigidities. As the Runge-Kutta method is computationally more expensive and and does not conserve particle energy, we have adopted the Boris-Push method for all the results in this paper.

\subsection{Derivative vs fractional definition of running diffusion coefficients\label{sec:deriv_frac}}

\begin{figure}[tbh]
\centering
\includegraphics[scale=1]{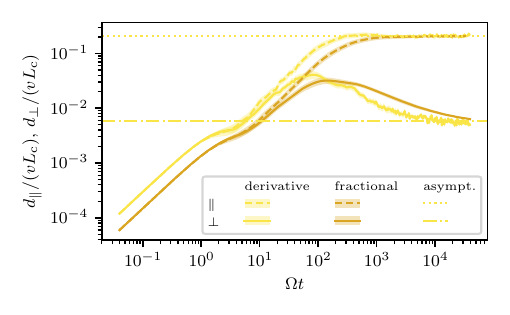}
\caption{Comparison of the running diffusion coefficients obtained with the derivative and fractional definitions, see eqs.~\eqref{eqn:def_dperp} and \eqref{eqn:def_dperp_frac}.}
\label{fig:dpar_dperp_deriv_frac}
\end{figure}

Finally, another systematic error in the running diffusion coefficients is due to the use of different definitions. We have used the derivative definitions, see eqs.~\eqref{eqn:def_dpar} and \eqref{eqn:def_dperp} throughout. However, in the literature, oftentimes the fractional definition, see eqs.~\eqref{eqn:def_dpar_frac} and \eqref{eqn:def_dperp_frac} is employed. 

In Fig.~\ref{fig:dpar_dperp_deriv_frac}, we compare the results for one combination of input parameters, that is $\rg/\Lc = 8.9 \times 10^{-3}$ and $\eta = 0.5$. At early times, that is in the ballistic phase, the running diffusion coefficients following from the derivative and fractional definitions differ by about a factor two as expected for mean-square displacements that are $\propto t^2$. At late times, that is in the diffusive limit where the mean-square displacements are $\propto t$, the definitions agree, of course. However, note that the convergence to the diffusive limit takes place later and more gradually for the fractional definition compared to the derivative definition. In practice, this means that the code needs to be run for longer times to achieve convergence of the running diffusion coefficients. While the nuisance numerical noise at late times is suppressed for the fractional definition, some of the interesting features at intermediate times are also less pronounced. Finally, the slope in the running perpendicular diffusion coefficient at intermediate times also differ. 

In summary, the derivative definition of the diffusion coefficient, besides being the one that allows to relate more closely to the TGK formalism also offers faster convergence and allows identifying features at intermediate times.

\end{document}